\documentclass[ reprint, aps, pra, showkeys, nofootinbib,floatfix]{revtex4-1}

\usepackage{amsfonts}
\usepackage{amssymb}
\usepackage{amsmath}
\usepackage{graphicx}
\usepackage{subfigure}
\usepackage{dcolumn}
\usepackage{bm}
\usepackage{color}
\usepackage{setspace}
\usepackage[hang]{footmisc}
\usepackage[sort&compress]{natbib}
\usepackage{multirow}
\usepackage[sort&compress]{cleveref}
\usepackage{nomencl}
\usepackage[english]{babel}
\usepackage{xcolor, soul}
\usepackage{marginnote}

\graphicspath{{./},{./figures/},{./graphs/},{./figure.Julian/}}
\sethlcolor{white}
\crefname{equation}{\unskip}{\unskip}
\crefname{figure}{\unskip}{\unskip}
\crefname{section}{\unskip}{\unskip}
\crefname{subsection}{\unskip}{\unskip}

\begin{document}

\title{Influence of anisotropic surface roughness on lubricated rubber friction with application to hydraulic seals}

\author{M. Scaraggi$^{1,2}$, J. Angerhausen$^{3}$,  L. Dorogin$^{2}$, H. Murrenhoff$^{3}$, B.N.J. Persson$^{1,4}$}
\affiliation{$^{1}$DII, Universit\'a del Salento, 73100 Monteroni-Lecce, Italy, EU}
\affiliation{$^{2}$PGI-1, FZ J\"ulich, Germany, EU}
\affiliation{$^{3}$IFAS, RWTH University, Germany, EU}
\affiliation{$^{4}$www.MultiscaleConsulting.com}

\date{\today }

\begin{abstract}
Machine elements and mechanical components have often surfaces with anisotropic roughness, which may result from the machining processes, e.g. grinding, or from wear. Hence, it is important to understand how surface roughness anisotropy affects contact mechanics properties, such as friction and the interface separation, which is important for lubricated contacts. Here we extend and apply a multiscale mean-field model to the lubricated contact between a soft (e.g. rubber) elastic solid and a rigid countersurface. We consider surfaces with anisotropic surface roughness, and discuss how the fluid flow factors and friction factors depend on the roughness. We present an experimental study of the lubricated sliding contact between a nitrile butadiene rubber O-ring and steel surfaces with different types of isotropic and anisotropic surface roughness. The good quantitative comparison between the experimental results and the theory predictions suggests that the multiscale lubrication mechanisms are accurately captured by the theory.
\end{abstract}

\maketitle
\makenomenclature


%

{\bf 1 Introduction}


Soft contacts are crucial in numerous engineering and biological
applications, and have gained increasing attention by scientists working in
fundamental science\cite%
{general_multiscale,Ma20157366,10.1371/journal.pone.0143415} and
computational mechanics\cite{Stupkiewicz2016511}. Soft contacts occur, e.g.,
in the tribology of machine elements\cite%
{Salant1999189,Hajjam200413,Wohlers200951,Scaraggi2012,Tan2015236,Scaraggi2014118}%
, in living matter\cite{Dunn201345,Khosla201517587,Greene20115255}, and
medical devices\cite{Lorenz2013,Sterner2016}, to name just a few. In some specific applications, the wet grip performance
becomes the subject of international regulation, such as for tires (e.g. the
EU tire label directive\cite{EU.tires}) and for personal protective shoes
(see e.g the test method for slip resistance of footwear ISO 13287:2012,
leading to the GRIP rating documents by the Health \& Safety Laboratory HSL,
UK). In some other applications, instead, different surface roughness descriptors
become the key parameters to be associated with the device failure behavior,
such as the International Roughness Index (IRI) for road pavement
engineering (shown to be related to the vehicles crash-rate\cite{pavement})
or a set of shaft roughness descriptors for rotatory seals\cite{Flitney} (to
be kept within an allowable range of values, see e.g. RMA OS-1-1 and ISO
6194-1:2009).

Interestingly, despite the social, economic and environmental relevancy of
the applications described above, the underlying tribological nature of the
generic soft contact has not been widely recognized. Consequently, a
strongly diverse design approach, often of the try-and-error nature, is used
in those apparently-different (but fundamentally similar) scientific and
technological fields. This might suggest that a dedicated applied research
is probably needed to inject the current theoretical tribology understanding
to practical applications of soft contacts.

A soft contact typically refers to an interaction between solids where at
least one of the solids is characterized by a relatively small elastic
modulus (say in the range $\mathrm{kPa}$ to $\mathrm{MPa}$). This results in
relative low contact pressures (typically $\sim 1 \ \mathrm{MPa}$ or less)
which imply that the fluid rheology can be treated as pressure independent,
i.e., the classical ``high" pressure lubricant behaviors, such as
piezo-viscosity and piezo-density, are unimportant. However, depending on
the contact conditions other complex phenomena may appears such as the
dewetting transitions (regulated by the interface spreading pressure). On
the theoretical side, the fundamental understanding of the key role of
roughness in soft interface mechanics, at least in those cases where
adhesion/capillary, hydration and weeping mechanisms can be neglected, has
been the subject of numerous investigations, in particular about the
multiscale description of real surfaces under dry \cite%
{Persson20013840,Carbone2009,general_multiscale} and wet \cite%
{BP,Scaraggi2014118,Scaraggi2012,MS1,MS2,MS3,MS4,MS5,MS6} conditions. In
Refs. \cite{BP,MS2,MS4} Persson, Scaraggi et al. have derived a mean-field
theory of the thin film lubrication, characterized by the so called flow and
shear stress factors\cite{Patir197812}. The theory can be applied to the
lubricated contact between soft solids with realistic surface roughness, for
both Newtonian and non-Newtonian fluids, and for anisotropic roughness.

In this work we will extend and apply the theory \cite{MS2} to practical
cases, trying to unravel how specific anisotropy roughness parameters
affect the so-called flow and stress factors which enter the fluid flow equation and
the expression for the frictional shear stress. A case study related to
dynamic seals with line contact geometry, such as O-rings, will be reported,
together with an experimental study. Indeed, among the others, seals are
crucial machine elements in almost any mechatronic device, and in particular
in hydraulic and pneumatic components, where a seal failure can result in
expensive production downtime or even environmentally hazardous leakage.

\begin{figure}[tbp]
\includegraphics[width=1.0\columnwidth]{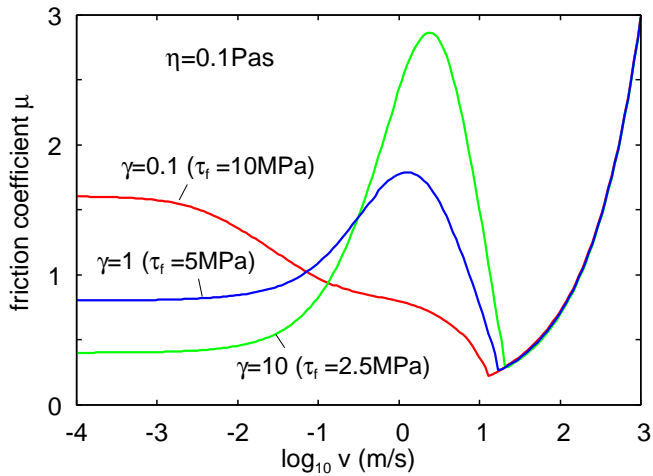}
\caption{The calculated friction coefficient (Stribeck curves) as a function of the logarithm of the sliding speed
        for three different surfaces with the same angular averaged power spectra but different surface asymmetry.
        For a rubber cylinder (radius $R=2.5 \ {\rm mm}$) with the normal load per unit length $F_{\rm N}/L = 200 \ {\rm N/m}$,
        sliding in the $x$-direction, orthogonal to the cylinder axis ($y$-direction).
        The substrate surface is rigid and randomly rough with the
Tripp numbers: $\gamma =1$ (isotropic roughness), $\gamma = 10$ (scratches along the sliding direction),
and $\gamma = 0.1$ (scratches orthogonal to the sliding direction). The rubber elastic modulus $E=5 \ {\rm MPa}$ and Poisson
ratio $\nu = 0.5$. The lubricant fluid is Newtonian with the viscosity $\eta = 0.1 \ {\rm Pas}$.
We assume in the area of real contact occur a frictional shear stress which equal $\tau_{\rm f} = 10$, $5$ and $2.5 \ {\rm MPa}$ for
$\gamma = 0.1$, $1$ and $10$, respectively. The substrate surfaces have the root-mean-square (rms) roughness $h_{\rm rms} = 10 \ {\rm \mu m}$, the
rms slope 1.5 and the Hurst exponent $H=0.8$.}
\label{1logv.2muTOT.eps}
\end{figure}

Surface roughness anisotropy can have different influence on the friction coefficient (and on wear)
depending on the sliding speed. To illustrate this, consider a rubber cylinder sliding orthogonal to the cylinder axis
on a lubricated substrate. The substrate is rigid and we consider two cases where the substrate has
roughness scratches orthogonal to the sliding direction i.e. along the cylinder axis (corresponding to
the Tripp number  $\gamma < 1$), and when the scratches are along the sliding direction (corresponding to
the Tripp number  $\gamma > 1$). We assume the same angular averaged surface roughness power spectra in both cases.

In the boundary lubrication velocity region the friction force will be highest 
when the scratches are orthogonal to the sliding direction since in this case
the rubber will be exposed to stronger time dependent deformations by the substrate asperities, resulting
in a large viscoelastic contribution to the friction.
For dry surfaces this has indeed been observed in experiments\cite{Carbone2009}.

In the high-velocity part of the mixed lubrication region of the Stribeck curve the situation is opposite to that in the boundary lubrication regime.
Thus when the scratches are orthogonal to the sliding direction, fluid in the valley cannot escape from the 
nominal rubber-substrate contact region as easily as when the scratches are along.
This results in a stronger hydrodynamic pressure buildup, and larger surface separation, and lower friction force.
This is illustrated in Fig. \ref{1logv.2muTOT.eps} where we show the calculated Stribeck curves 
(using the theory developed in Ref. \cite{MS2}) for three surfaces with
identical angular averaged surface roughness power spectra, but different anisotropy: $\gamma = 1$ (isotropic roughness),
$\gamma = 0.1$ (scratches along the cylinder axis, orthogonal to the sliding direction) and $\gamma = 10$ (scratches along the sliding direction).


The paper is organized as following. In Sec. 2 we provide a summary of the
mean field theory for lubrication, with emphasis on the theory of flow and
shear stress factors for the specific contact conditions matching the
experimental setup. Furthermore we present numerical results which
illustrate how different roughness parameters affect the flow and friction
factors, and the related friction curves. In Sec. 3 we describe the
experimental apparatus built for the friction tests, and in Sec. 4 we
present the surface roughness power spectra of the studied surfaces.
Sec. 5 reports the friction measurements. The measured data are compared in detail to theory
predictions in Sec. 6. Sec. 7 contains the summary and conclusions.

\vskip 0.5cm \textbf{2 Mean field theory of lubrication} 
\label{theory}

\begin{figure}[tbp]
\begin{center}
\includegraphics[width=0.8\columnwidth]{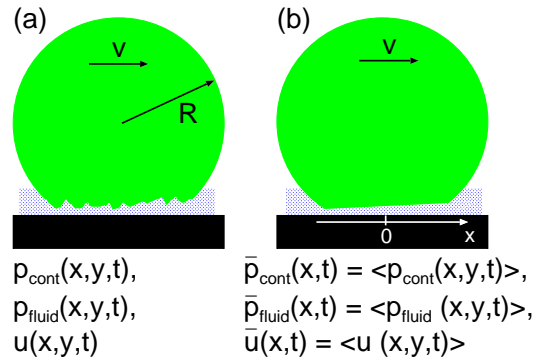}
\end{center}
\caption{(a) Schematic of a rubber ball with a rough surface sliding on a
smooth rigid substrate surface. Physical quantities like the contact
pressure, the fluid pressure and the interface separation varies rapidly in
space over many decades in length scales due to the nature of the surface
roughness. The complex situation in (a) can be mapped on a simpler situation
(b) where the fluid and contact pressures, and the surface separation, are
locally-averaged quantities, which vary slowly in space on the length scale
of the surface roughness. Those averaged quantities obey to modified fluid
flow equations which contain two functions, denoted as flow factors, which
depend on the locally averaged surface separation, and which are mainly
determined by the surface roughness.}
\label{twoBALL.eps}
\end{figure}

\vskip 0.2cm \textbf{2.1 Effective equations of motion}

We consider the simplest problem of an elastic cylinder (upper solid with
length $B$ and radius $R$, with $B>>R$) sliding on a rigid solid (lower
solid) with a nominal flat surface. We consider two cases:

(A) The upper solid has surface roughness while the lower solid is perfectly
smooth.

(B) The upper solid is perfectly smooth while the lower solid has surface
roughness.

We assume that the sliding occurs in the direction perpendicular to the
cylinder axis, and we introduce a coordinate system with the $x$-axis along
the sliding direction and with $x=0$ corresponding to the cylinder axis
position, see the schematic of Fig. \ref{twoBALL.eps}. The cylinder is
squeezed against the substrate by the normal force $F_{\mathrm{N}}$. In the
contact region between the cylinder and the substrate acts a nominal contact
pressure: 
\begin{equation*}
\bar{p}_{0}(x,t)=\bar{p}_{\mathrm{cont}}(x,t)+\bar{p}_{\mathrm{fluid}}(x,t),%
\eqno(1)
\end{equation*}%
where $\bar{p}_{\mathrm{cont}}$ and $\bar{p}_{\mathrm{fluid}}$ are,
respectively, the mean field pressures due to the direct asperity- and
fluid-asperity interactions. Note that $\bar{p}_{0}(x,t)$ is the microscopic
pressure averaged over surface areas $\sim \lambda \times \lambda $, where $%
\lambda $ is the wavelength of the longest (relevant) surface roughness
component. This approach assumes a separation of length scales so that $%
\lambda <<w$, where $w$ is the width of the nominal contact region (which is
of order the Hertz's contact length $a_{\mathrm{H}}$). We consider the case
of constant normal load so that 
\begin{equation*}
\int_{-\infty }^{\infty }dx\ \bar{p}_{0}(x,t)={\frac{F_{\mathrm{N}} }{B}}.%
\eqno(2)
\end{equation*}

Let $\bar{u}(x,t)$ denote the (locally averaged) separation between the
surfaces. For $\bar{u}>h_{\mathrm{rms}}$, where $h_{\mathrm{rms}}$ is the
root-mean-square (rms) roughness,%
\begin{equation*}
\bar{p}_{\mathrm{cont}}(x,t)\approx \beta E_{\mathrm{r}}\mathrm{exp}\left(
-\alpha {\frac{\bar{u}(x,t)}{h_{\mathrm{rms}}}}\right) ,\eqno(3)
\end{equation*}%
where $\alpha $ and $\beta $ are described in Ref. \cite{Persson2007}. Eq.
(3) is valid for large enough $\bar{u}$. Since an infinite high pressure is
necessary in order to squeeze the solids into complete contact we must have $%
\bar{p}_{\mathrm{cont}}\rightarrow \infty $ as $\bar{u}\rightarrow 0$. This
is, of course, not obeyed by (3), and in the following calculations we
therefore use the numerically calculated relation $\bar{p}_{\mathrm{cont}}(%
\bar{u})$, the latter reducing to (3) for large enough $\bar{u}$.

The macroscopic gap equation is determined by simple geometrical
considerations. Thus, assuming the cylinder deformation to be within the
Hertz regime for elastic solids, the gap equation reads 
\begin{equation*}
\bar{u}(x,t)=\bar{u}_{0}(t)+{\frac{x^{2}}{2R}}-{\frac{2}{\pi E_{\mathrm{r}}}}%
\int_{-\infty }^{\infty }dx^{\prime }\ \bar{p}_{0}(x^{\prime },t)\mathrm{ln}%
\left\vert {\frac{x-x^{\prime }}{x^{\prime }}}\right\vert ,\eqno(4)
\end{equation*}%
where in the most general line-contact case $1/R=1/R^{\mathrm{up}}+1/R^{%
\mathrm{low}}$, where $R^{\mathrm{up}}$ and $R^{\mathrm{low}}$ are the
radius of curvature of the top and bottom surface, respectively. In (3) and
(4) $1/E_{\mathrm{r}}=1/E_{\mathrm{r}}^{\mathrm{up}}+1/E_{\mathrm{r}}^{%
\mathrm{low}}$ in the most general case, where the reduced elastic modulus
is defined by $E_{\mathrm{r}}=E/\left( 1-\nu ^{2}\right) $, with $E$ and $%
\nu $ the Young's elastic modulus and the Poisson ratio, respectively. In
addition, the pressure $\bar p_{0}(x,t)$ must satisfy the normal load
conservation condition (2).

To complete the system of equations we need an equation which determine the
fluid pressure $\bar{p}_{\mathrm{fluid}}(x,t)$. The fluid flow is
determined by the Navier-Stokes equation, but in the present case of fluid
flow in a narrow gap between the solid walls, this equation can be
simplified under the so called lubrication approximation, leading to the
Reynolds equation. For surfaces with roughness on many length scales, this
equation (even reproducing only the laminar motion), when coupled with (4),
is numerically too complex to be solved directly in most cases. However,
when there is a separation of length scales, i.e., if the longest (relevant)
surface roughness component has a wavelength much smaller than the nominal
cylinder-countersurface contact length (in the sliding direction), then it
is possible to eliminate or (in the language of the renormalization group
theory) integrate out the surface roughness degrees of freedom and obtain an
effective equation of motion for the (locally averaged) fluid pressure. Such
equations are characterized by two correction factors, namely $\phi _{%
\mathrm{p}}$ (pressure flow factor) and $\phi _{\mathrm{s}}$ (shear flow
factor), which are mainly determined by the surface roughness and depend on
the locally averaged surface separation $\bar{u}$. Thus, the effective
two-dimensional (2D) fluid flow current for the case A (stationary cylinder
with roughness)\cite{MS2}: 
\begin{equation*}
\mathbf{J}=-{\frac{\bar{u}^{3}\phi _{\mathrm{p}}(\bar{u})}{12\eta }}\nabla 
\bar{p}_{\mathrm{fluid}}+{\frac{1}{2}}\bar{u}\mathbf{v}_{0}+{\frac{1}{2}}h_{%
\mathrm{rms}}\phi _{\mathrm{s}}(\bar{u})\mathbf{v}_{0}\eqno(5)
\end{equation*}%
satisfies the mass conservation equation 
\begin{equation*}
{\frac{\partial \bar{u}}{\partial t}}+\nabla \cdot \mathbf{J}=0\eqno(6)
\end{equation*}%
Substituting (5) in (6), and assuming $\mathbf{v}=v_{0}\hat{x}$, gives the
modified Reynolds equation: 
\begin{equation*}
{\frac{\partial \bar{u}}{\partial t}}={\frac{\partial }{\partial x}}\left[ {%
\frac{\bar{u}^{3}\phi _{\mathrm{p}}(\bar{u})}{12\eta }}{\frac{\partial \bar{p%
}_{\mathrm{fluid}}}{\partial x}}-{\frac{1}{2}}\bar{u}v_{0}-{\frac{1}{2}}h_{%
\mathrm{rms}}\phi _{\mathrm{s}}(\bar{u})v_{0}\right] .\eqno(7)
\end{equation*}%
The set of equations (1) to (4), and (7), represents 5 equations for the 5
unknown variables $\bar{p}_{0}$, $\bar{p}_{\mathrm{cont}}$, $\bar{p}_{%
\mathrm{fluid}}$, $\bar{u}$ and $\bar{u}_{0}$. Finally, considering steady
sliding, $\partial \bar{u}/\partial t=0$, (7) is solved with Cauchy boundary
conditions, whereas the cavitation Reynolds boundary condition is set at
macroscopic scale by requiring\footnote{%
We note that for soft elastic solids, like rubber, we have shown in Ref. 
\cite{PS0} that including cavitation or not has no drastic effect on the
result, and in particular the friction coefficient as a function of the
sliding speed, $\mu =\mu (v)$, is nearly unchanged.} $\bar{p}_{\mathrm{fluid}%
}\geq 0$ .

It is useful to define an effective viscosity $\eta_{\mathrm{eff}} = \eta /
\phi_{\mathrm{p}}$ which depends on the surface roughness via $\phi_{\mathrm{%
p}}$. When studying fluid squeeze-out with $v_0 = 0$ (no sliding), the
pressure flow factor (or the effective viscosity $\eta_{\mathrm{eff}}$) is
the only parameter in the fluid flow dynamics (effective Reynolds equation)
where the roughness enter.

\vskip 0.2cm \textbf{2.2 Symmetry properties of the pressure flow factors}

The flow current (5) (case A) was derived in Ref. \cite{MS2} assuming that
the upper solid is stationary with surface roughness, while the substrate
was assumed perfectly flat (no roughness) and moving with the velocity $%
\mathbf{v}_{0}$. For this case we will present numerical results below. We
will also compare the theory with experimental results for another case
(case B) where the upper solid is assumed to be perfectly smooth, whereas
the substrate has roughness. The flow current for this situation is again
given by (5) but with the shear flow factor replaced by $\phi _{\mathrm{s}%
}\rightarrow -\phi _{\mathrm{s}}$, while $\phi _{\mathrm{p}}$ is unchanged.
This result can be proved by considering the mapping shown in Fig. \ref%
{ShearFlowFactorInvert1.eps}.

\begin{figure}[tbp]
\includegraphics[width=0.75\columnwidth]{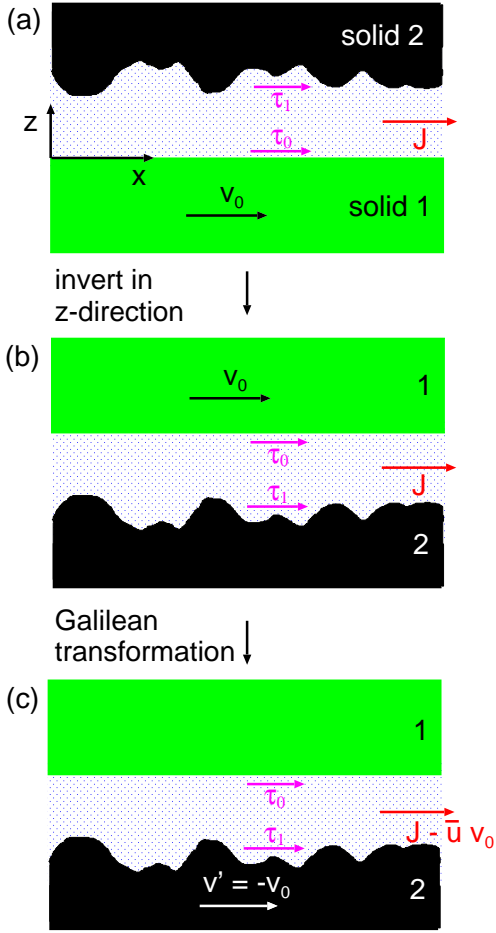}
\caption{Mapping used to show how the flow factors change when the surface
roughness of the two surfaces is interchanged.}
\label{ShearFlowFactorInvert1.eps}
\end{figure}

We first invert the system along the $z$-direction. This leaves the flow
current unchanged. Next we make a Galilean transformation to a new reference
frame where the upper solid is stationary and the lower solid moves with the
velocity $\mathbf{v}^{\prime }= - \mathbf{v}_0$. In the new reference frame
the flow current equals $\mathbf{J}_{\mathrm{new}} = \mathbf{J} - \bar u 
\mathbf{v}_0$. For the original sliding system (case A; see Fig. \ref%
{ShearFlowFactorInvert1.eps}(a)) the average fluid flow current is given by
(5) so that for the final system (case B, Fig. \ref%
{ShearFlowFactorInvert1.eps}): 
\begin{equation*}
\mathbf{J}_{\mathrm{new}}=\mathbf{J}-\bar{u}\mathbf{v}_{0}
\end{equation*}%
\begin{equation*}
=-{\frac{\bar{u}^{3}\phi _{\mathrm{p}}(\bar{u})}{12\eta }}\nabla \bar{p}_{%
\mathrm{fluid}}+{\frac{1}{2}}\bar{u}\mathbf{v}_{0}+{\frac{1}{2}}h_{\mathrm{%
rms}}\phi _{\mathrm{s}}(\bar{u})\mathbf{v}_{0}-\bar{u}\mathbf{v}_{0}
\end{equation*}%
\begin{equation*}
=-{\frac{\bar{u}^{3}\phi _{\mathrm{p}}(\bar{u})}{12\eta }}\nabla \bar{p}_{%
\mathrm{fluid}}-{\frac{1}{2}}\bar{u}\mathbf{v}_{0}+{\frac{1}{2}}h_{\mathrm{%
rms}}\phi _{\mathrm{s}}(\bar{u})\mathbf{v}_{0}
\end{equation*}%
\begin{equation*}
=-{\frac{\bar{u}^{3}\phi _{\mathrm{p}}(\bar{u})}{12\eta }}\nabla \bar{p}_{%
\mathrm{fluid}}+{\frac{1}{2}}\bar{u}\mathbf{v}^{\prime }-{\frac{1}{2}}h_{%
\mathrm{rms}}\phi _{\mathrm{s}}(\bar{u})\mathbf{v}^{\prime }.
\end{equation*}%
Hence, denoting $\mathbf{v}^{\prime }$ with $\mathbf{v}_{0}$ for simplicity,
we get 
\begin{equation*}
\mathbf{J}_{\mathrm{new}}=-{\frac{\bar{u}^{3}\phi _{\mathrm{p}}(\bar{u})}{%
12\eta }}\nabla \bar{p}_{\mathrm{fluid}}+{\frac{1}{2}}\bar{u}\mathbf{v}_{0}-{%
\frac{1}{2}}h_{\mathrm{rms}}\phi _{\mathrm{s}}(\bar{u})\mathbf{v}_{0}.\eqno%
(8)
\end{equation*}

Thus the flow current for case B (the sliding system (c) in Fig. \ref%
{ShearFlowFactorInvert1.eps}) can be obtained from that of case A (the
original sliding system (a) in Fig. \ref{ShearFlowFactorInvert1.eps}) by
replacing $\phi _{\mathrm{s}}$ with $-\phi _{\mathrm{s}}$, while $\phi _{%
\mathrm{p}}$ is unchanged.


\vskip 0.2cm \textbf{2.3 The frictional shear stress}

Let us now study the frictional stress acting on the solids at the
interface. We are interested in the locally averaged (effective) frictional
stress $\bar{\tau}_{\mathrm{f}}=\langle \tau _{\mathrm{f}}(\mathbf{x}%
)\rangle $. This is obtained by averaging $\tau _{\mathrm{f}}(\mathbf{x})$
over small surface areas of size $\sim \lambda \times \lambda $, where $%
\lambda $ is the wavelength of the longest (relevant) surface roughness
component. The effective frictional stress has a contribution from the area
of real contact and another from the fluid in the non-contact area, and we
write 
\begin{equation*}
\bar{\tau}_{\mathrm{f}}=\bar{\tau}_{\mathrm{cont}}+\bar{\tau}_{\mathrm{fluid}%
}.\eqno(9)
\end{equation*}%
The (nominal) frictional stress resulting from the area of solid-solid
contact acting on the lower surface (we assume the lower surface to be
moving with the velocity $v_{0}$ while the upper solid is stationary) is
derived in Appendix A: 
\begin{gather*}
\bar{\tau}_{\mathrm{cont}}^{\mathrm{up}}(x,t)=\tau _{1}{\frac{A_{1}(x,t)}{%
A_{0}}} -\bar{p}_{\mathrm{cont}}(x,t) \times \\ \left [ {E_{\rm r}^{\rm up} \over  E_{\rm r}^{\rm up}+E_{\rm r}^{\rm down}}\nabla \bar{u}^{\mathrm{up}}(x,t)+
{E_{\rm r}^{\rm down} \over E_{\rm r}^{\rm up}+E_{\rm r}^{\rm down}}
\nabla \bar{u}^{\rm down}(x,t) \right ] ,
\end{gather*}%
where the shear stress $\tau _{1}$ is positive when the sliding direction is
along the positive $x$-direction, and negative otherwise. Note that $%
A_{1}(x,t)=A(x,t,\zeta _{1})$ is the area of real contact at the highest
magnification, and $A_{0}$ the nominal contact area (of the elementary
volume of roughness). $\bar{u}^{\mathrm{down}}(x,t)$ and $\bar{u}^{\mathrm{up%
}}(x,t)$ are the locally averaged (deformed) lower and upper surface, with $%
\bar{u}(x,t)=\bar{u}^{\mathrm{up}}(x,t)-\bar{u}^{\mathrm{down}}(x,t)$. The
term proportional to $\bar{p}_{\mathrm{cont}}$ is the solid
macro-rolling friction contribution acting on the upper solid, coming from
the projection, onto the sliding direction, of the effective solid contact
elementary forces. A similar friction stress (but with opposite sign on the
adhesion contribution) acts on the upper solid: 
\begin{gather*}
\bar{\tau}_{\mathrm{cont}}^{\mathrm{down}}(x,t)=-\bar{\tau}_{\mathrm{cont}}^{\mathrm{up}}(x,t)
\end{gather*}

In this paper we consider only the line contact case, and in this case all
mean field quantities, e.g., the local area of contact $A(x,y)$, depends
only on $x$, as already assumed above. Furthermore, in the present
calculations we neglect the adhesion between the solids. This is a valid
assumption for positive spreading pressures, in which case there will be an
effective short-ranged repulsion between the surfaces. Here we will also
assume that the shear stress $\tau _{1}$ is independent of the local
pressure and of the sliding speed. For most materials, e.g., rubber, $\tau
_{1}$ will depend on the sliding speed but to illustrate the effect of the
fluid dynamics on the friction, we prefer to keep $\tau _{1}$ as a constant. 

\begin{figure}[tbp]
\includegraphics[width=0.75\columnwidth]{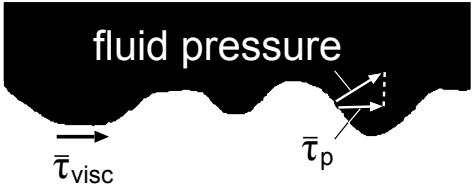} 
\caption{The shear stress acting on the solid walls from the fluid has a
contribution $\protect\tau_{\mathrm{visc}}$ from shearing the viscous fluid
(proportional to the fluid viscosity) and a second contribution $\protect\tau%
_{\mathrm{p}}$ from the tangential component of the fluid pressure $p(x,y)$
acting on the solid wall.}
\label{pictureBothStress.eps}
\end{figure}

The effective frictional shear stress $\bar{\tau}_{\mathrm{fluid}}$
originating from the fluid has, in general, two 
contributions (see Fig. \ref{pictureBothStress.eps}):
\begin{equation*}
\bar{\tau}_{\mathrm{fluid}}=\bar{\tau}_{\mathrm{visc}}+\bar{\tau}_{\mathrm{p}%
} .
\end{equation*}%
The first term $\bar{\tau}_{\mathrm{%
visc}}$ is determined by the effective shear stresses acting on the solid
walls, whereas a second term $\bar{\tau}_{\mathrm{p}}$ is due to the
projection of the effective fluid pressure forces along the sliding
direction, i.e., $\bar{\tau}_{\mathrm{p}}^{\mathrm{down}}=\bar{p}_{\mathrm{%
fluid}}(x,t)\nabla {\bar{u}}^{\mathrm{down}}(x,t)$ for the lower surface,
and ${\bar{\tau}}_{\mathrm{p}}^{\mathrm{up}}=-\bar{p}_{\mathrm{fluid}%
}(x,t)\nabla {\bar{u}}^{\mathrm{up}}(x,t)$ for the upper surface, where ${%
\bar{u}}^{\mathrm{down}}(x,t)=\left\langle u^{\mathrm{down}%
}(x,y,t)\right\rangle $ and ${\bar{u}}^{\mathrm{up}}(x,t)=\left\langle u^{%
\mathrm{up}}(x,y,t)\right\rangle $ are the locally-averaged (deformed)
surface of the bottom and upper solid, respectively (as before). Thus, $\bar{\tau}_{%
\mathrm{visc}}$ takes into account the rough nature of the contact in the
determination of the local effective frictional stresses, and in particular
we show in the following that $\bar{\tau}_{\mathrm{visc}}$ includes a
contribution coming from the projection of the local (asperity-scale) fluid
pressure forces along the sliding direction (which we refer to as a
micro-rolling friction term). The fluid shear stress is given by%
\begin{equation*}
\tau _{\mathrm{visc}}=\eta {\frac{\partial v_{x}}{\partial z}},\eqno(11)
\end{equation*}%
and on the bottom (down) and upper (up) surfaces it reads, respectively (all
quantities depends on time but this variable is here and in what follows
suppressed for convenience), 
\begin{equation*}
\tau _{\mathrm{visc}}^{\mathrm{down}}\left( \mathbf{x}\right) =-\frac{\eta 
\mathbf{v}_{0}}{u(\mathbf{x})}-{\frac{1}{2}}u(\mathbf{x})\nabla p(\mathbf{x}%
),\eqno(12a)
\end{equation*}%
\begin{equation*}
\tau _{\mathrm{visc}}^{\mathrm{up}}\left( \mathbf{x}\right) =\frac{\eta 
\mathbf{v}_{0}}{u(\mathbf{x})}-{\frac{1}{2}}u(\mathbf{x})\nabla p(\mathbf{x}%
),\eqno(12b)
\end{equation*}%
resulting in%
\begin{equation*}
\tau _{\mathrm{fluid}}^{\mathrm{down}}\left( \mathbf{x}\right) =\tau _{%
\mathrm{visc}}^{\mathrm{down}}\left( \mathbf{x}\right) +p(\mathbf{x})\nabla
u^{\mathrm{down}}(\mathbf{x}),\eqno(13a)
\end{equation*}%
\begin{equation*}
\tau _{\mathrm{fluid}}^{\mathrm{up}}\left( \mathbf{x}\right) =\tau _{\mathrm{%
visc}}^{\mathrm{up}}\left( \mathbf{x}\right) -p\left( \mathbf{x}%
\right)\nabla u^{\mathrm{up}}(\mathbf{x}).\eqno(13b)
\end{equation*}%
Here $p=p_{\mathrm{fluid}}$ for simplicity of notation.

We will now calculate $\bar{\tau} _{\mathrm{fluid}}$ for one case only,
namely when the upper solid is elastic and has random surface roughness and
the lower solid is rigid and with a flat surface (no surface roughness). The
other cases of interest can be studied in a similar way, and the derivation
and results are given in Appendix B.

\textbf{For the bottom surface}: Since the bottom surface is flat and rigid, 
$\nabla u^{\mathrm{down}}=0$ and hence $\tau _{\mathrm{fluid}}^{\mathrm{down}%
}=\tau _{\mathrm{visc}}^{\mathrm{down}}$. Using (12a) we get in this case
(see also Ref. \cite{MS2}): 
\begin{equation*}
\langle \tau _{\mathrm{visc}}\rangle =\bar{\tau}_{\mathrm{visc}}=-\phi _{%
\mathrm{f}}\frac{\eta_0 \mathbf{v}_{0}}{\bar{u}}-{\frac{1}{2}}\left\langle
u\nabla p\right\rangle ,\eqno(14)
\end{equation*}%
where\cite{MS2} 
\begin{equation*}
\phi _{\mathrm{f}}=\left\langle {\frac{\eta }{u(\mathbf{x})}}\right\rangle {%
\frac{\bar{u}}{\eta _{0}}},\eqno(15)
\end{equation*}%
where $\eta _{0}$ is the low shear rate fluid viscosity. Furthermore, for $%
\bar{u}<\bar{u}_{\mathrm{c}}$ (where $\bar{u}_{\mathrm{c}}$ is the average
interfacial separation at the point when the contact area percolate) 
\begin{equation*}
\langle u\nabla p\rangle =6\eta _{0}\mathbf{v}_{0}\bar{u}\left\langle {\frac{%
1}{u}}\right\rangle .\eqno(16)
\end{equation*}
\begin{figure*}[tbp]
\includegraphics[width=1.8\columnwidth]{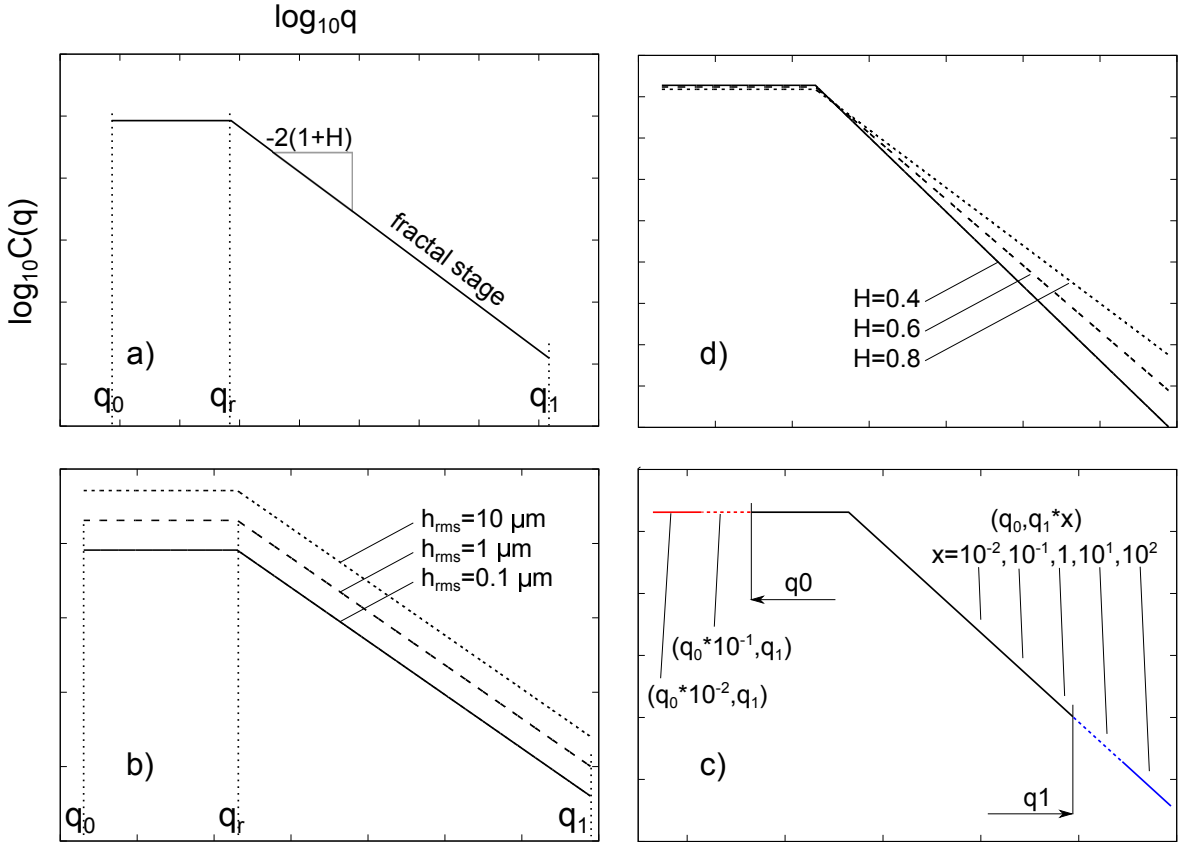}
\caption{Schematic of the power spectral density (PSD) adopted to elucidate
the role of roughness properties on the flow and friction factors. (a) The
reference surface is characterized by a fractal content in the range of
frequencies $\left[ q_{\mathrm{r}},q_{1}\right] $, and a flat (roll-off) PSD
in $\left[ q_{0},q_{\mathrm{r}}\right] $. The latter PSD range provides a
simplified mathematical representation of the typical decaying PSD behavior,
at increasing wavenumbers, shown by real surfaces. (b) The comparison is
made at constant wavenumbers and fractal dimension, but at varying rms 
surface roughness $h_{\mathrm{rms}}$. (c) Here $h_{\mathrm{rms}}$ and
the fractal dimension are kept constant, whereas the wavenumbers $q_0$ and $%
q_1$ are alternately varied. Finally, (d) both wavenumbers $q_0$ and $q_1$,
and $h_{\mathrm{rms}}$ are kept fixed, but the Hurst exponent $H$ ($H=3-D_{%
\mathrm{f}}$, where $D_{\mathrm{f}}$ is the fractal dimension) is varied.}
\label{image.PSD.eps}
\end{figure*}
On the other hand for $\bar{u}>>h_{\mathrm{rms}}$ one has \cite{MS2} 
\begin{equation*}
\langle u\nabla p\rangle =\left( 1-3D{\frac{\langle h^{2}\rangle }{\bar{u}%
^{2}}}\right) \bar{u}\nabla \bar{p}+6D{\frac{\langle h^{2}\rangle }{\bar{u}%
^{2}}\frac{\eta _{0}\mathbf{v}_{0}}{\bar{u}},}\eqno(17)
\end{equation*}%
where $\bar{p}=p_{\mathrm{fluid}}$. The general form for (16) and (17) is 
\begin{equation*}
\langle u\nabla p\rangle =\phi _{\mathrm{fp}}\bar{u}\nabla \bar{p}+\phi _{%
\mathrm{fs}}\frac{2\mathbf{v}_{0}\eta _{0}}{\bar{u}},\eqno(18)
\end{equation*}%
where the relations for $\phi _{\mathrm{fp}}$ and $\phi _{%
\mathrm{fs}}$, obtained by interpolation, are given in Ref. \cite{MS2}: 
\begin{equation*}
\phi _{\mathrm{fs}}={\frac{3\bar{u}}{\langle u^{-1}\rangle ^{-1}+\theta (%
\bar{u}-\bar{u}_{\mathrm{c}})(\bar{u}-\bar{u}_{\mathrm{c}})(\bar{u}%
^{2}/\langle h^{2}\rangle )D^{-1}}},\eqno(19)
\end{equation*}%
and 
\begin{equation*}
\phi _{\mathrm{fp}}={\frac{\bar{u}(\bar{u}-\bar{u}_{\mathrm{c}})\theta (\bar{%
u}-\bar{u}_{\mathrm{c}})}{\bar{u}^{2}+3\langle h^{2}\rangle D}},\eqno(20)
\end{equation*}%
where $\theta (u)$ is the unit step function and the surface roughness
anisotropic parameter, $D$, is related to the Tripp number $\gamma $ with $%
D=\left( 1+\gamma \right) ^{-1}$ \cite{MS2}. Substituting (18) in (14) gives
the effective frictional shear stress acting on the bottom solid 
\begin{equation*}
\bar{\tau}_{\mathrm{visc}}^{\mathrm{down}}=-\left( \phi _{\mathrm{f}}+\phi _{%
\mathrm{fs}}\right) {\frac{\eta _{0}\mathbf{v}_{0}}{\bar{u}}}-\frac{1}{2}%
\phi _{\mathrm{fp}}\bar{u}\nabla \bar{p}.\eqno(21)
\end{equation*}%
where $\phi _{\mathrm{f}}$, $\phi _{\mathrm{fs}}$ and $\phi _{\mathrm{fp}}$
are given by (15), (19) and (20).

\textbf{For the top surface}: The top surface has surface roughness and the
fluid rolling term 
\begin{equation*}
p({\bf x})\nabla u({\bf x}),\eqno(22)
\end{equation*}%
is non-vanishing, where $u^{\mathrm{up}}({\bf x})=u({\bf x})$ in this case. Hence on
the top surface the fluid acts with the local friction shear stress 
\begin{equation*}
\tau _{\mathrm{fluid}}^{\mathrm{up}}=\tau _{\mathrm{visc}}^{\mathrm{up}%
}-p\left( \mathbf{x}\right) \nabla u\left( \mathbf{x}\right) .\eqno(23)
\end{equation*}%
From (13) and (23) we get 
\begin{equation*}
\langle \tau _{\mathrm{fluid}}^{\mathrm{up}}\rangle =\bar{\tau}_{\mathrm{%
fluid}}^{\mathrm{up}}=\left\langle {\frac{\eta \mathbf{v}_{0}}{u}}%
\right\rangle -{\frac{1}{2}}\langle u\nabla p\rangle -\langle p\nabla
u\rangle .\eqno(24)
\end{equation*}%
To second order in the rms-roughness amplitude one has (see Ref. \cite{MS2}%
): 
\begin{equation*}
\left\langle p\nabla u\right\rangle =\left\langle p_{1}\nabla
u_{1}\right\rangle +\bar{p}\nabla \bar{u},\eqno(25)
\end{equation*}%
and 
\begin{equation*}
\left\langle u\nabla p\right\rangle =\left\langle u_{1}\nabla
p_{1}\right\rangle +\bar{u}\nabla \bar{p}.\eqno(26)
\end{equation*}%
Next, using Eqs. (A5)-(A9) in Ref. \cite{MS2} it is easily shown that 
\begin{equation*}
\left\langle p_{1}\nabla u_{1}\right\rangle =-\left\langle u_{1}\nabla
p_{1}\right\rangle .\eqno(27)
\end{equation*}%
Using (25)-(27) we get 
\begin{equation*}
\left\langle p\nabla u\right\rangle =-\left\langle u\nabla p\right\rangle +%
\bar{p}\nabla \bar{u}+\bar{u}\nabla \bar{p}.\eqno(28)
\end{equation*}%

Substituting (28) in (24) gives 
\begin{equation*}
\bar{\tau}_{\mathrm{fluid}}^{\mathrm{up}}=\left\langle \frac{\eta }{u}%
\right\rangle \mathbf{v}_{0}+{\frac{1}{2}}\left\langle u\nabla
p\right\rangle -\bar{u}\nabla \bar{p}-\bar{p}\nabla \bar{u},\eqno(29)
\end{equation*}%
The last term in (29), $-\bar{p}\nabla \bar{u}$, is the fluid macro-rolling
frictional shear stress $\bar{\tau}_{\mathrm{p}}^{\mathrm{up}}$ (since $-%
\bar{p}\nabla \bar{u}^{\mathrm{up}}=-\bar{p}\nabla \bar{u}$, given the
bottom rigid and flat surface). Using (15), (18) and (29) we get 
\begin{equation*}
\bar{\tau}_{\mathrm{fluid}}^{\mathrm{up}}=\left( \phi _{\mathrm{f}}+\phi _{%
\mathrm{fs}}\right) {\frac{\eta _{0}\mathbf{v}_{0}}{\bar{u}}}+{\frac{1}{2}}%
(\phi _{\mathrm{fp}}-2)\bar{u}\nabla \bar{p}-\bar{p}\nabla \bar{u}.\eqno(30)
\end{equation*}%
We do note that the global mechanical equilibrium along the sliding
direction is conserved, i.e. 
\begin{equation*}
\int_{x_{a}}^{x_{b}}dx~\bar{\tau}_{\mathrm{fluid}}^{\mathrm{down}%
}+\int_{x_{a}}^{x_{b}}dx~\bar{\tau}_{\mathrm{fluid}}^{\mathrm{up}}+\bar{p}_{%
\mathrm{out}}\bar{u}_{\mathrm{out}}-\bar{p}_{\mathrm{in}}\bar{u}_{\mathrm{in}%
}=0,\eqno(31)
\end{equation*}%
where $\bar{p}_{\mathrm{out}}$ ($\bar{p}_{\mathrm{in}}$) and $\bar{u}_{%
\mathrm{out}}$ ($\bar{u}_{\mathrm{in}}$) are, respectively, the average
fluid pressure and separation at the contact outlet (inlet).

\vskip0.2cm \textbf{2.4 Dependence of flow and friction factors on roughness
parameters}

In this section we present numerical results for the fluid flow factors $%
\phi _{\mathrm{p}}$ and $\phi _{\mathrm{s}}$, and for the frictional stress
factors $\phi _{\mathrm{f}}$, $\phi _{\mathrm{fs}}$ and $\phi _{\mathrm{fp}}$%
, for solids with different surface roughness. Note that the accurate 
description of the flow and shear stress factors, and of the
asperity-asperity contact, is crucial in the formulation of any mean field
lubrication theory.

The power spectral densities (PSD) $C(q)$ adopted in this section are shown
in Fig. \ref{image.PSD.eps}. Fig. \ref{image.PSD.eps}(a) shows the PSD of a
surface with a (self affine) fractal region between the roll-off wavenumber $%
q_{\mathrm{r}}$ and the large wavenumber cut-off $q_{1}$. Most surfaces of
engineering interest have a nearly flat roll-off region $q_{0}<q<q_{\mathrm{r%
}}$, where $C(q) $ is almost constant\cite{Persson2005}. Fig. \ref{image.PSD.eps}(b)
shows three power spectra with the same $q_{0}$, $q_{\mathrm{r}}$ and $q_{1}$
wavenumbers, and the same fractal dimension, but with different rms 
surface roughness $h_{\mathrm{rms}}$. Fig. \ref{image.PSD.eps}(c) shows the
case where $h_{\mathrm{rms}}$ and the fractal dimension are kept constant,
whereas the large and small wavenumber cut-off are varied. Finally, in Fig. %
\ref{image.PSD.eps}(d) both the cut-off and roll-off wavenumbers and $h_{%
\mathrm{rms}}$ are fixed, but the Hurst exponent $H$ ($H=3-D_{\mathrm{f}}$,
where $D_{\mathrm{f}}$ is the fractal dimension) is varied.

\begin{figure}[tbh]
\centering%
\includegraphics[
                        width=0.9\columnwidth,
                        angle=0
                ]{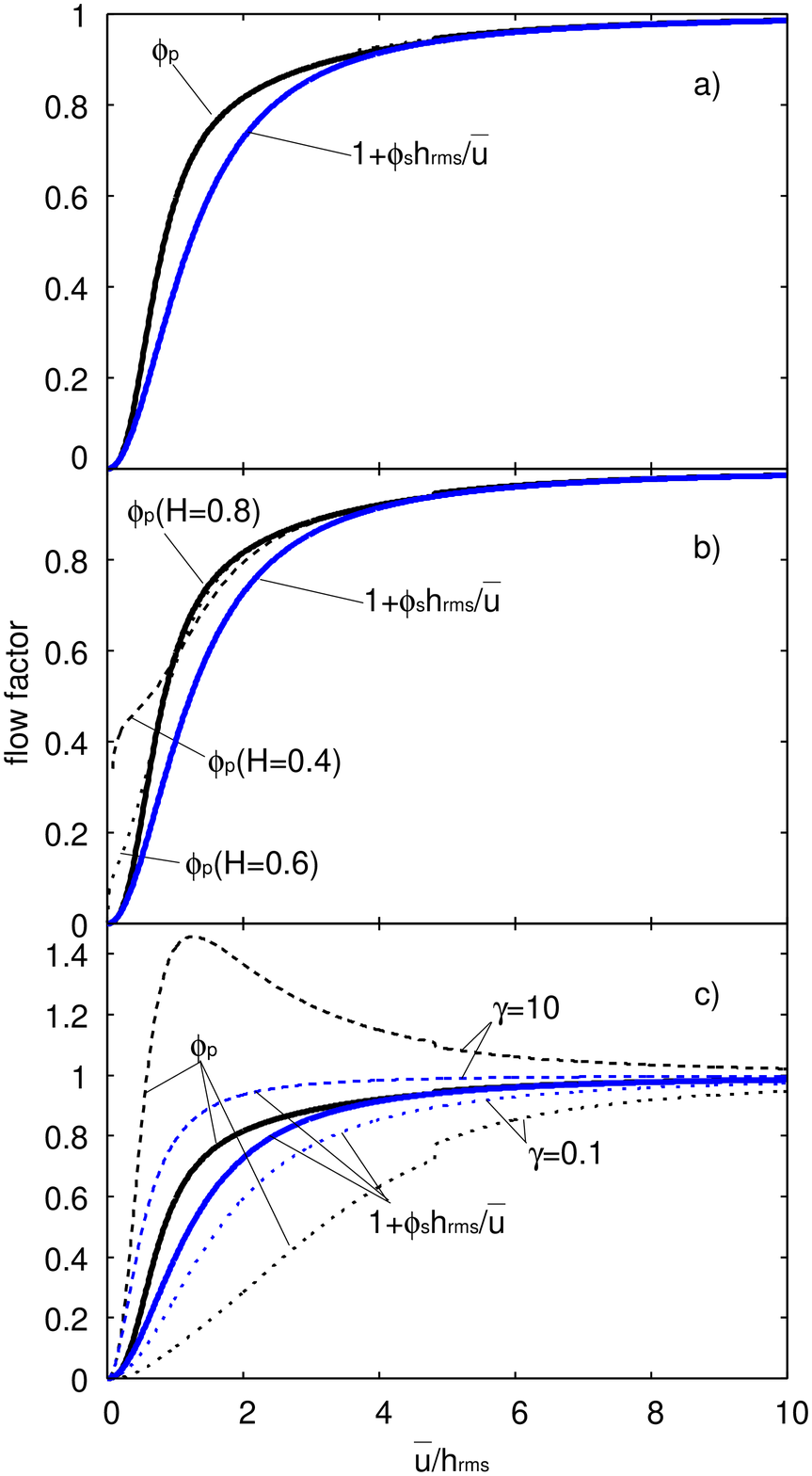}
\caption{The mean field correction to the Poiseuille flow $\protect\phi _{%
\mathrm{p}}$ (black lines) and the mean field correction to the Couette flow 
$\left( 1+\protect\phi _{\mathrm{s}}h_{\mathrm{rms}}/\bar{u}\right) $\ (blue
lines), as a function of the dimensionless average interface separation $%
\bar{u}/h_{\mathrm{rms}}$. In particular, in (a) we vary $h_{\mathrm{rms}}$,
the small wavenumber cut-off $q_{0}$ and the large wavenumber cut-off $q_{1}$%
, whereas in (b) we vary the Hurst exponent $H$ and in (c) we vary the Tripp
number.}
\label{flow.factor.A.eps}
\end{figure}

The reference PSD has the small wavenumber cut-off $q_{0}=2\times 10^{3}~%
\mathrm{m}^{-1}$, the roll-off wavenumber $q_{r}=2\times 10^{5}~\mathrm{m}%
^{-1}$, the large wavenumber cut-off $q_{1}=7.8\times 10^{9}~\mathrm{m}^{-1}$%
, the rms roughness $h_{\mathrm{rms}}=1 ~\mathrm{\mu m}$
and the Hurst exponent $H=0.8$. The reference PSD correspond to a surface
with the rms-slope $2.45$, so the surface with the rms-roughness $10 \ 
\mathrm{\mu m}$, and the surfaces with the Hurst exponent $H=0.6$ and $0.4$
will have too large rms-slope to be physically reasonable, but for the
present parameter study this is not very important. In what follows we
assume that the elastic solid has the reduced Young's modulus $E_{\mathrm{r}%
}=6.67~\mathrm{MPa}$, and the fluid is assumed to be Newtonian.

In Fig. \ref{flow.factor.A.eps} we show the mean field correction to the
Poiseuille flow $\phi _{\mathrm{p}}$ (black lines) and the mean field
correction to the Couette flow $\left( 1+\phi _{\mathrm{s}}h_{\mathrm{rms}}/%
\bar{u}\right) $\ (blue lines), as a function of the dimensionless average
interface separation $\bar{u}/h_{\mathrm{rms}}$. In Fig. \ref%
{flow.factor.A.eps}(a) we vary $h_{\mathrm{rms}}$, the small wavenumber
cut-off $q_{0}$ and the large wavenumber cut-off $q_{1}$, whereas in (b) we
vary the Hurst exponent $H$ and in (c) we vary the Tripp number. In Fig. \ref%
{flow.factor.A.eps}(a) all curves superpose, i.e., the flow factors as a
function of $\bar u /h_{\mathrm{rms}}$ are not sensitive to $q_{0}$, $q_{1}$
and $h_{\mathrm{rms}}$, at least not in the parameter range studied here.
This is, however, not the case when we vary the Hurst exponent $H$ (see Fig. %
\ref{flow.factor.A.eps}(b)) or the Tripp number $\gamma$ (see Fig. \ref%
{flow.factor.A.eps}(c)). Let us explain for the pressure flow factor $\phi_{%
\mathrm{p}}$ the physical origin of this.

When we decrease the Hurst exponent, the roughness at short length scales
increases, while the long wavelength roughness is nearly unchanged (assuming
a fixed $h_{\mathrm{rms}}$ roughness amplitude). This implies that when $H$
decreases the surface rms-slope increases, which in turn decreases the
contact area. Thus the pressure needed for the contact area to percolate (at
which point $\phi_{\mathrm{p}}$ vanish) will increase when $H$ decreases.
Since increasing pressure implies decreasing average interfacial separation
it follows that the value of $\bar u /h_{\mathrm{rms}}$
where $\phi_{\mathrm{p}}$ first vanish will decrease as $H$ decreases, in
agreement with Fig. \ref{flow.factor.A.eps}(b).
Here we have used that the average interfacial separation is 
mainly determined by the long wavelength roughness and therefore only weakly
dependent on the Hurst exponent, assuming a constant $h_{\mathrm{rms}}$.

Fig. \ref{flow.factor.A.eps}(c) shows that when we increase the Tripp number
the fluid pressure flow factor increases. Since the pressure flow factor
defines an effective viscosity $\eta_{\mathrm{eff}} = \eta /\phi_{\mathrm{p}}
$ (see Sec. 2.1), this is equivalent to the statement that an increasing
Tripp number result in a lower effective viscosity. This is easy to
understand since a large Tripp number $\gamma >> 1$ implies that there are
roughness channels (in our case grinding scratches) along the fluid flow
direction (i.e., orthogonal to the cylinder axis), which facilitate the flow
of the fluid out from the nominal contact area. In the opposite limit $%
\gamma << 1$ the flow channels are orthogonal to the fluid flow direction
(i.e. along the cylinder axis) which inhibit fluid squeeze-out and result in
an effective viscosity which is larger than the bare viscosity for all
interfacial separations.

\begin{figure}[tbh]
\centering
\includegraphics[
                        width=0.9\columnwidth,
                        angle=0
                ]{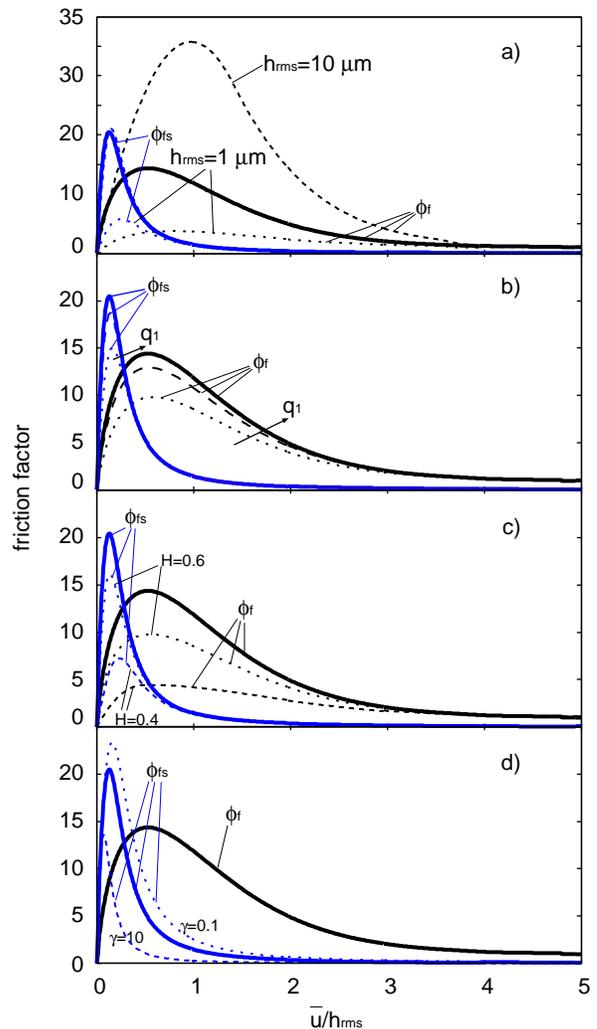}
\caption{The mean field correction to the sliding shear stress ($\protect%
\eta v_{0}/\bar{u}$) $\protect\phi _{\mathrm{f}}$ (black lines) and $\protect%
\phi _{\mathrm{fs}}$ (blue lines), as a function of the dimensionless
average interface separation $\bar{u}/h_{\mathrm{rms}}$. In (a) we vary $h_{\mathrm{rms}}$
and the small wavenumber cut-off $q_{0}$, in (b) the large wavenumber cut-off $q_{1}$,
in (c) the dependence on the fractal
dimension and in (d) on the Tripp number.}
\label{fi.ffs.dep}
\end{figure}

\begin{figure}[tbh]
\centering
\includegraphics[
                        width=0.5\columnwidth,
                        angle=0
                ]{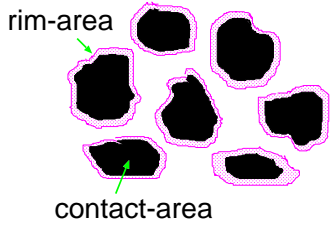}
\caption{ The contact area (black) and the rim area (pink). The most
important contribution to $\langle u^{-1} \rangle$ comes from the pink area,
i.e., from relative narrow strips of surface area at the rim of the contact
area, where the surface separation is very small. }
\label{ContactRimArea.eps}
\end{figure}

\begin{figure}[tbh]
\centering
\includegraphics[
                        width=0.8\columnwidth,
                        angle=0
                ]{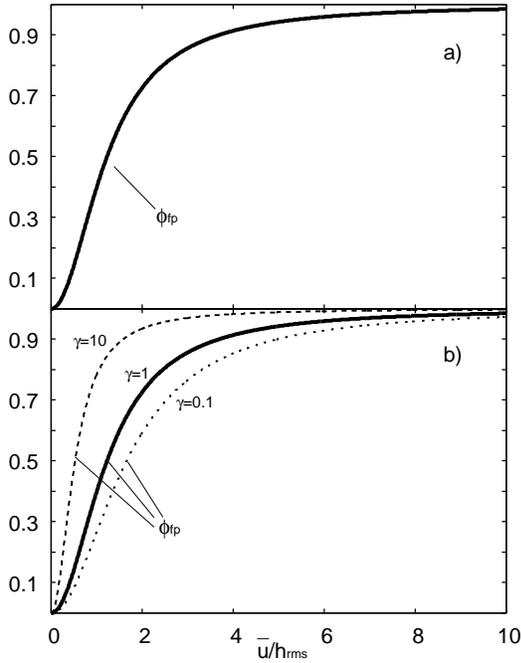}
\caption{The mean field correction to the pressure gradient friction term ($%
\bar{u}\protect\nabla p$) $\protect\phi _{\mathrm{fp}}$ as a function of the
dimensionless average interface separation $\bar{u}/h_{\mathrm{rms}}$. In
particular, in (a) we vary $h_{\mathrm{rms}}$, the small wavenumber cut-off $%
q_{0}$ and the large wavenumber cut-off $q_{1}$ as well as the Hurst
exponent $H$, whereas in (b) we vary the Tripp number.}
\label{fi.ffp.dep}
\end{figure}

In Fig. \ref{fi.ffs.dep} we show the mean field correction $\phi _{\mathrm{f}%
}$ (black lines) to the sliding shear stress ($\eta v_{0}/\bar{u}$), and $%
\phi _{\mathrm{fs}}$ (blue lines), as a function of the dimensionless
average interface separation $\bar{u}/h_{\mathrm{rms}}$. In Fig. \ref%
{fi.ffs.dep}(a) we show the dependence on $h_{\mathrm{rms}}$ and on the small
wavenumber cut-off $q_{0}$. In Fig. \ref{fi.ffs.dep}(b) we vary the
large wavenumber cut-off $q_{1}$,
in Fig. \ref{fi.ffs.dep}(c) the fractal dimension, and in
Fig. \ref{fi.ffs.dep}(d) the Tripp number. 
We first remark that $%
q_{0}$ has a negligible influence on the sliding shear stress
(curves are superposed), whereas $q_1$, $h_{\mathrm{rms}}$ and $D_{\mathrm{f}}$ do
influence the sliding shear stresses.

Fig. \ref{fi.ffs.dep}(a) shows that $\phi _{\mathrm{f}}$ increases with
increasing $h_{\mathrm{rms}}$. This can be understood from the definition $%
\phi _{\mathrm{f}}\sim \left\langle \bar{u}/u\right\rangle =\bar u \langle
u^{-1} \rangle$. Now, as $h_{\mathrm{rms}}$ becomes larger $\bar u$ will
increase too [at low nominal contact pressures it is nearly proportional to $%
h_{\mathrm{rms}}$ (see \cite{Persson2007})]. The quantity $\langle u^{-1} \rangle$ has its
biggest contribution from the surface area where the separation $u$ is very
small. This surface area is located in a narrow rim around the area of real
contact (see Fig. \ref{ContactRimArea.eps}). We will denote the non-contact
surface area, where the surface separation is below, say $u< 10 \ \mathrm{nm}
$, as the rim area. If $h_{\mathrm{rms}}$ is increased by adding only
long-wavelength roughness to the surface profile then the rms-slope, and
hence also the area of real contact, and the rim area, are nearly unchanged.
In this case $\langle u^{-1} \rangle$ will be nearly unchanged, and $\phi _{%
\mathrm{f}}$ will increase with increasing $h_{\mathrm{rms}}$. However, in
the present case the surface roughness height is scaled everywhere by a
multiplicative factor, say $\zeta$ [equal to 10 or 0.1 in Fig. \ref%
{fi.ffs.dep}(a)]. In this case, at a constant applied pressure, if $\zeta$
increases, $\bar u$ will increase with roughly a factor of $\zeta$, while the
contact area will decrease with roughly a factor $1/\zeta$. However, it is
likely that the rim area will decrease slower with $\zeta$ then the surface
area since it depends on the length of the boundary lines of the contact
area which may decrease slower than the area itself. This would explain why $%
\phi _{\mathrm{f}}$, as a function of $\bar u /h_{\rm rms}$, 
increases as $h_{\mathrm{rms}}$ increases.

Similar considerations apply to $\phi _{\mathrm{fs}}$ in the range of
average interface separation $\bar{u}<u_{\mathrm{c}}$, where again $\phi _{%
\mathrm{fs}}\sim \left\langle \bar{u}/u\right\rangle $. At larger
separations, instead, all $\phi _{\mathrm{fs}}$ curves in Fig. \ref%
{fi.ffs.dep}(a) converge to a unique mastercurve, being $\phi _{\mathrm{fs}%
}\sim {h_{\mathrm{rms}}^{2}/}\bar{u}^{2}$.

Fig. \ref{fi.ffs.dep}(b) shows that when the large wavenumber cut-off $q_1$ 
increases, $\phi_{\rm f}$ increases too. This can be understood as follows: 
Note that $\bar u$ is mainly determined by the long wavelength roughness
and hence does not depend on the cut-off $q_1$ unless $\bar u/ h_{\rm rms} << 1$.
We now consider the change in the contact as we move from the cut-off $q_1'=0.1q_1$
to $q_1$. We consider first the system with the largest wavenumber cut-off and write $q=\zeta q_0$
with $q_1 = \zeta_1 q_0$. 
When we increase the magnification $\zeta$ the contact area
decreases, i.e., new non-contact area appears in what appeared to be contact
at a lower magnification. When the magnification is increased to $\zeta=0.1 \zeta_1$
we observe what is essentially the contact as it would prevail for the 
system with the cut-off $q'_1=0.1 q_1$.
When we increase the magnification from $\zeta=0.1 \zeta_1$ to $\zeta=\zeta_1$
some regions which appeared to be in contact at the magnification $\zeta=0.1 \zeta_1$,
will now be observed to be non-contact regions with closely spaced surfaces, while the non-contact regions
observed at the magnification $\zeta=0.1 \zeta_1$ will still be non-contact regions
with nearly the same surface separation as observed at the lower 
magnification. The ``additional'' non-contact area (with closely spaced surfaces) 
observed when increasing the magnification from $\zeta=0.1 \zeta_1$ to $\zeta=\zeta_1$
will result in an important additional contribution to 
$\langle u^{-1} \rangle$. We conclude that $\phi_{\rm f}$ as a function
of $\bar u /h_{\rm rms}$ will increase as $q_1$ increases.

When the fractal dimension increases (or Hurst exponent $H$ decreases), both 
$\phi _{\mathrm{f}}$ and $\phi _{\mathrm{fs}}$ decreases as shown in Fig. %
\ref{fi.ffs.dep}(c). This is easy to understand. When $H$ decreases the
amplitude of the short wavelength roughness increases, while the long
wavelength roughness, assuming a constant $h_{\mathrm{rms}}$, change very
little. When the short wavelength roughness increases, the surface rms-slope
and hence the surface area decreases. This imply that the rim area and hence
also $\langle u^{-1} \rangle$ decreases as the Hurst exponent $H$ decreases.
It follows that $\phi _{\mathrm{f}}$ will decrease when the fractal
dimension increases at constant $h_{\mathrm{rms}}$.

Fig. \ref{fi.ffs.dep}(d) shows that $\phi _{\mathrm{f}}$ is unaffected by
the anisotropy (Tripp number $\gamma $) of the surface. This is indeed
expected because both the contact area (and the rim area) and the average
surface separation are nearly independent of the surface anisotropy
assuming identical angular averaged surface roughness power spectrum.
Indeed, this hold exactly within the Persson's contact mechanics theory
where only the angular-averaged power spectral density \cite{Carbone2009}
enters. As a consequence, $\phi _{\mathrm{f}}$ is unaffected by the Tripp
number. The situation is different in the case of $\phi _{\mathrm{fs}}$,
which at larger separation $\phi _{\mathrm{fs}}\sim D$, where $D=\left(
1+\gamma \right) ^{-1}$.

Finally, in Fig. \ref{fi.ffp.dep} we show $\phi _{\mathrm{fp}}$, which is
the mean field correction to the pressure gradient friction term, $\bar{u}%
\nabla p$, as a function of the dimensionless average interface separation $%
\bar{u}/h_{\mathrm{rms}}$. In Fig. \ref{fi.ffp.dep}(a) we vary $h_{\mathrm{%
rms}}$, the small wavenumber cut-off $q_{0}$ and the large wavenumber
cut-off $q_{1}$ as well as the Hurst exponent $H$, whereas in Fig. \ref%
{fi.ffp.dep}(b) we vary the Tripp number. Here $\phi _{\mathrm{fp}}$ is
mainly sensitive to the Tripp number through its dependence with $D $.
However, differently from $\phi _{\mathrm{fs}}$, $\phi _{\mathrm{fp}}$
increases at increasing Tripp numbers. Thus, given the opposite behaviors
shown by the various frictional sources ($\phi _{\mathrm{f}}$, $\phi _{%
\mathrm{fs}}$ and $\phi _{\mathrm{fp}}$) originating from the fluid viscous
dissipation in anisotropically rough contacts, it is not possible to
establish a general statement about the role of the Tripp number in wet
contact mechanics, since the total friction will be determined by the
relative weight and the variation of such friction factors terms along the
effective contact domain. However, qualitatively, a mixed lubrication
occurring under $\eta v_{0}/\left\vert \nabla p\right\vert \gg \bar{u}^{2}$,
thus at relatively large values of contact pressures (leading to a
small average interface separation) will be characterized by larger friction
for transverse roughness than for longitudinal roughness, in agreement with
experimental findings \cite{MS7}. For $\bar{u}^{2}\gg \eta v_{0}/\left\vert
\nabla p\right\vert $, instead, the contact is expected to occur under small
true contact area, leading to large interfacial separation and thus to
larger friction for the longitudinal roughness than for the transverse
roughness. The latter case will be discussed in the comparison with
experimental results provided in the following.

\begin{figure}[tbh]
\centering%
\includegraphics[
                        width=1.0\columnwidth,
                        angle=0
                ]{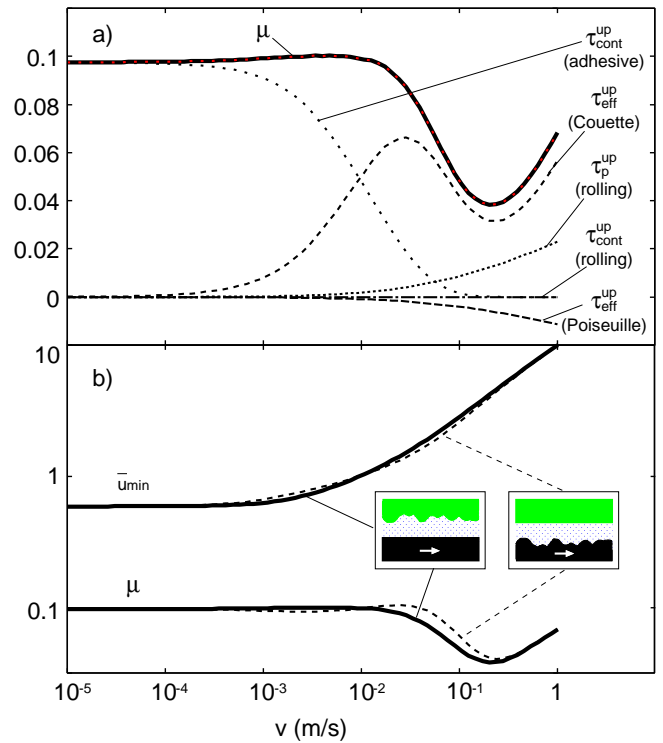}
\caption{(a) The friction (black solid line) and the corresponding dissipation
terms (other black lines) originating on the cylinder, as a function of the
sliding speed. The solid black line is for the case of a rough steady
cylinder of radius $R=2.5~\mathrm{mm}$ squeezed against a flat ideally
smooth rigid sliding counter surface, under constant normal line load $F_{\mathrm{N%
}}=100~\mathrm{N/m}$. we consider a solid with a reduced elastic modulus $E_{%
\mathrm{r}}=6.67~\mathrm{MPa}$, as well as a Newtonian fluid, with dynamic
viscosity $\protect\eta =0.1~\mathrm{Pas}$. The shear stress occurring in
the true contact areas is assumed constant and valued $\protect\sigma _{%
\mathrm{f}}=1~\mathrm{MPa}$. The sliding speed is in the range between $%
10^{-5}$ and $1~\mathrm{m/s}$. The red solid line, instead, is for the same
contact conditions of before but with the elasticity assigned to the bottom
sliding surface (i.e. the cylinder is rigid and the substrate is
deformable). (b) The friction coefficient (the lower solid and dashed lines) and
the minimum average separation (the upper solid and dashed lines), as a
function of the sliding speed, for the same contact condition of (a), but for the case of roughness on the cylinder (solid
lines) or on the substrate (dashed lines). The full and dashed lines are for
the cases where the cylinder is rigid and compliant, respectively.}
\label{friction.nominal}
\end{figure}
\begin{figure}[tbh]
\centering%
\includegraphics[
                        width=1.0\columnwidth,
                        angle=0
                ]{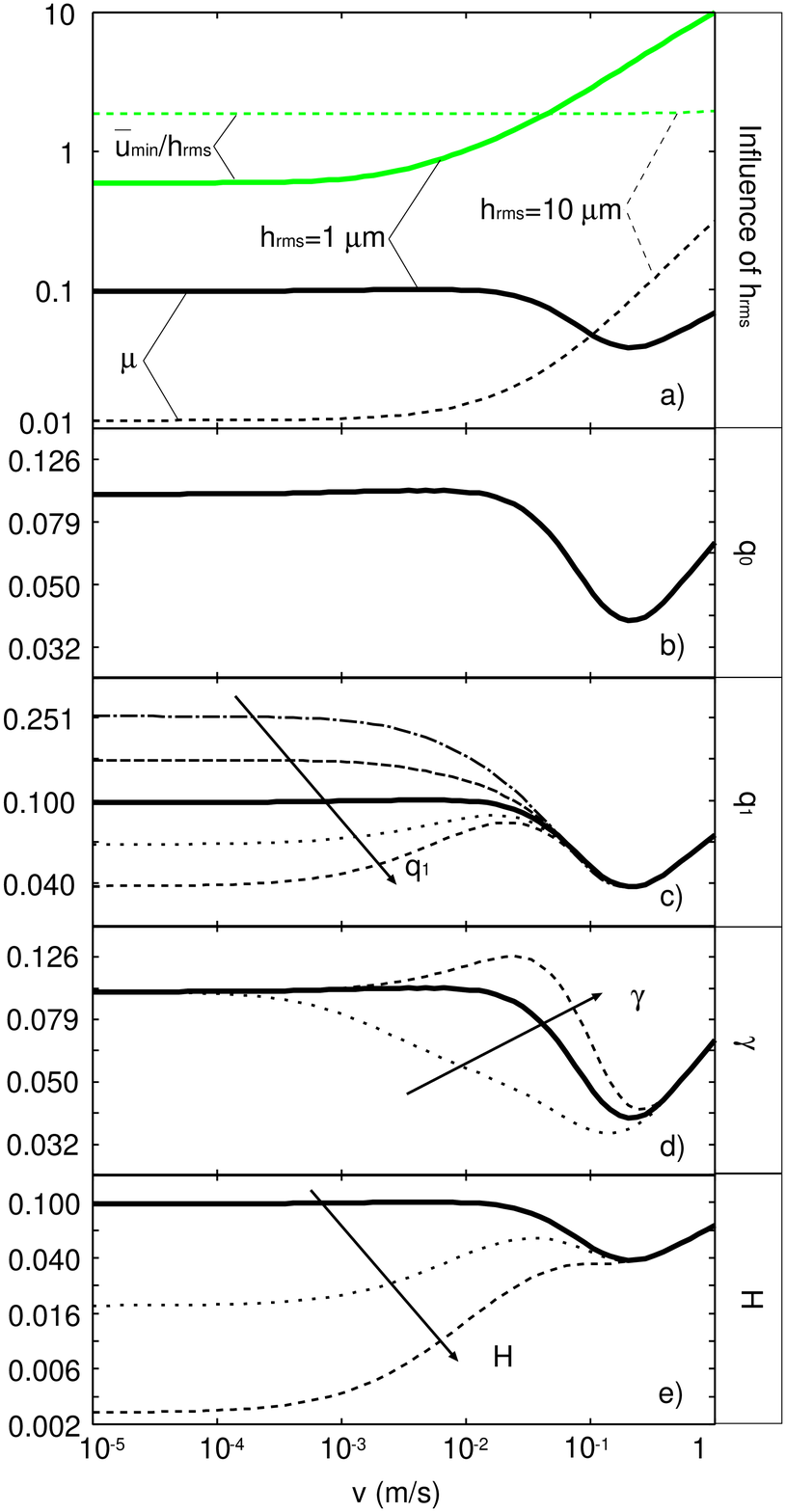}
\caption{The friction as a function of the sliding speed, for the same
parameters of Fig. \protect\ref{friction.nominal}(a) but for (from top to
bottom): (a) varying $h_\mathrm{rms}$, (b) $q_0$, (c) $q_1$, (d) $\protect%
\gamma$ and (e) Hurst exponent. The PSDs are described in Fig. \protect\ref%
{image.PSD.eps}.}
\label{friction.all}
\end{figure}

\vskip0.2cm \textbf{2.5 Dependence of lubricated friction on roughness
parameters}

In this section we present numerical results for the Stribeck curve, for
solids with different surface roughness. The power spectral densities (PSD) $%
C(q)$ adopted in this section are shown in Fig. \ref{image.PSD.eps}. Again
we consider a solid with reduced elastic modulus $E_{\mathrm{r}}=6.67~%
\mathrm{MPa}$, and assume a Newtonian fluid with the dynamic viscosity $\eta
=0.1~\mathrm{Pas}$. The flow and friction factors used in the calculation
below are those discussed before. In the following, unless differently
specified, we consider a stationary cylinder of radius $R=2.5~\mathrm{mm}$,
and with random surface roughness, squeezed against a flat ideally smooth
sliding (speed $v_0$) counter surface. On the cylinder we apply the normal (line
contact) load per unit length $F_{\mathrm{N}}/L=100~\mathrm{N/m}$, and the shear stress
occurring in the true contact areas is assumed constant $\tau_1 =1~\mathrm{%
MPa}$. The sliding speed is varied from $10^{-5}$ to $1~\mathrm{m/s}$.

We write the friction force acting on the cylinder as $F=\mu F_{\mathrm{N}}$%
. In Fig. \ref{friction.nominal}(a) we show the friction coefficient $\mu$
(solid line), and the different contributions to the friction coefficient
(dashed lines), as a function of the sliding speed. Thus, from top to bottom:

(1) the adhesive contribution to friction (i.e. the friction coming from the
shearing action occurring in true solid contact area), proportional to the
shear stress $\tau^{\mathrm{up}}_{\mathrm{cont}} (\mathrm{adhesive})$,

(2) the Couette-flow term [proportional to $\tau^{\mathrm{up}}_{\mathrm{eff}%
}(\mathrm{Couette})$],

(3) the rolling friction originating from the fluid pressure field
[proportional to $\tau^{\mathrm{up}}_{\mathrm{p}}(\mathrm{rolling})$],

(4) the rolling friction originating from the solid contact pressure field
[proportional to $\tau^{\mathrm{up}}_{\mathrm{cont}}(\mathrm{rolling})$],

(5) the Poiseuille-flow term [proportional to $\tau^{\mathrm{up}}_{\mathrm{%
eff}}(\mathrm{Poiseuille})$].

We note that the rolling friction coming from the solid contact pressure is
zero, since the substrate is nominally flat and rigid (and,anyway,
no solid bulk viscoelasticy is considered in this paper). In the same figure, the red solid line (approximately
superposed with the black dashed line) is for the same contact conditions of
before, but with the elasticity assigned to the bottom sliding surface (i.e.
the cylinder is rigid and the substrate is deformable). We note that for the
adopted set of contact parameters, the friction force is the same in
the two cases.

Let us now instead consider the case where the roughness is transferred from the
steady cylinder to the sliding substrate. Thus in Fig. \ref{friction.nominal}%
(b) we show the friction coefficient (the lower solid and dashed lines) and
the minimum average separation (the upper solid and dashed lines), as a
function of the sliding speed, for the same contact condition of Fig. \ref%
{friction.nominal}(a), but for the case of roughness on the cylinder (solid
lines) or on the substrate (dashed lines). We note,
as expected, that the different allocation of the roughness does play a role
only in the mixed lubrication regime, where the asperity-fluid interactions
dominate the contact mechanics.

In Fig. \ref{friction.all} we show the friction as a function of the sliding
speed (and in Fig. \ref{friction.all}(a) also the minimum surface
separation), for the same parameters as in Fig. \ref{friction.nominal}(a),
except that we vary (from top to bottom): (a) $h_\mathrm{rms}$, (b) $q_0$,
(c) $q_1$, (d) $\gamma$ and (e) the Hurst exponent. Fig. \ref{friction.all}%
(a) shows that when the rms roughness increases from $1\ \mathrm{\mu m}$
(solid green line) to $10\ \mathrm{\mu m}$ (dashed green line), the build up
of hydrodynamic fluid pressure with increasing sliding speed is not enough
to increase the average surface separation, i.e., the (nominal) fluid
pressure is negligible compared to the (nominal) asperity contact pressure.
Thus, for $h_{\mathrm{rms}}=10\ \mathrm{\mu m}$ the minimum interface
separation does not increase with increasing sliding speed, 
i.e. for all velocities studied it takes the same value as in the boundary lubrication
regime. Nevertheless, Fig. \ref{friction.all}(a) shows that there is a
hydrodynamic contribution to the friction force (black dashed line). This is
due to the shearing of the fluid in the small asperity gaps, and is
described by the $\phi_\mathrm{f}$ term in the expression for the frictional
shear stress. Therefore, in this case the hydrodynamic contribution to the
friction is a truly micro-EHL effect, i.e. a hydrodynamic effect occurring
at the scale of the roughness asperities.

In Fig. \ref{friction.all}(b) and (c), we show the effect of the large and
small roughness wavelength, respectively. As expected, $q_0$ has negligible
effect on the friction curve, because it has only very small effect on the
roughness solid contact mechanics. The large frequency cut-off $q_1$ (and
the Hurst exponent, see Fig. \ref{friction.all}(e)), instead, has a direct
influence on the solid contact mechanics, in particular through the mean
square roughness slope $\langle (\nabla h)^2 \rangle$, which increases when $%
q_1$ increases. Thus, since the true contact area $A_\mathrm{c}$ (in the
linear regime) scales as $A_\mathrm{c}\propto F_\mathrm{N} \langle (\nabla
h)^2 \rangle^{-1/2}$, the friction coefficient in the boundary regime will
be proportional to $\langle (\nabla h)^2 \rangle^{-1/2}$. Hence, the contact
area and the friction coefficient will decrease at increasing $q_1$ (and
decreasing $H$, see Fig. \ref{friction.all}(e)). Finally, in Fig. \ref%
{friction.all}(d) we show the effect of roughness anisotropy on the
friction. As anticipated, for very low sliding speed, when hydrodynamic
effects are negligible, the contact area, and hence also the friction force,
does not depend on the anisotropy parameter $\gamma$ (note: this is true only if there
is no viscoelastic contribution to the friction; 
see Fig. \ref{1logv.2muTOT.eps} and the discussion
related to this figure). However, at higher
sliding speeds (when $\bar{u}^{2}\gg \eta v_{0}/\left\vert \nabla
p\right\vert $), where hydrodynamic effects becomes important, when the long
axis of the (on the average elliptic) asperities are aligned along the
sliding direction, fluid is more easily removed from the nominal contact
region, resulting in smaller surface separation and larger friction force,
than when the asperities are aligned perpendicular to the sliding direction.

\vskip0.5cm \textbf{2.6 Some comments on mean-field lubrication models}

As pointed out above, an accurate treatment of
the fluid flow and friction shear stress factors, as well as of the
asperity-asperity contact, is crucial in the formulation of a mean
field lubrication theory. To this point we observe that, (at least) the
roughness contact mechanics has been the subject of dedicated and very
detailed research efforts in recent years, leading to some
controversial research papers but also to some incontestable results. Among
the latter outcomes we remark the recognition, by analytical\cite{general_multiscale}, numerical\cite{Scaraggi.in.prep,scaraggi2016effect} 
and experimental approaches\cite{lorenz2010leak,lorenz2013adhesion,lorenz2013static}, of the exponential
relation between the average contact pressure $p_{\mathrm{cont}}$ and
interface separation $\bar{u}$ in elastic contacts 
for large enough surface separation [see e.g. (3)]\cite{Persson2007}, as
well as of the linearity between true contact area and contact pressure (for 
$A(p)/A_0 < 0.3$) in both elastic and viscoelastic contacts\cite{Persson2014}, the
occurrence of solid percolation and its effects on the hydraulic resistivity
(see e.g. (16) and related discussion)\cite{lorenz2010leak}, and many others.

As an example, consider the contact pressure vs average separation relation, 
$\bar p=p(\bar u)$, as given by (3). This relation follows from the Persson
multiscale contact mechanics approach\cite{Persson2007} in the limit of $\bar{u}/h_{%
\mathrm{rms}}>>1 $, and is in agreement with both experimental\cite{0953-8984-21-1-015003} and
numerical studies\cite{Almqvist20112355}, except at very low nominal contact pressures where
finite size effects occurs\cite{Almqvist20112355}. Therefore, for elastic solids with randomly
rough surfaces any other pressure vs separation law (and related contact
mechanics theory) not showing such an exponential behavior (e.g., the GW
many-noninteracting-asperity theory\cite{Greenwood300}), should be avoided as building
block for a mean-field lubrication theory.

In a recent paper Masjedi and Khonsari \cite{Khonsari} observe that similar
results are obtained when adopting the GW contact mechanics, as obtained in
the 2009 paper by Persson and Scaraggi\cite{MS1}, where a mean field lubrication
theory was developed based on the Persson contact mechanics theory. Here we
note that in Ref. \cite{Khonsari} two of the three roughness parameters,
namely the mean asperity radius $\beta$ and the asperity surface density $n$%
, where used as fitting parameters. Indeed, the authors claim that such
parameters are difficult to obtain from a power spectral density \cite%
{Khonsari}, but in fact can be calculated as described by Nayak \cite%
{nayak}. In particular: 
\begin{equation*}
\frac{1}{\beta ^{2}}=\frac{8m_{4}}{3\pi}
\end{equation*}%
and 
\begin{equation*}
n=\frac{1}{6\pi \sqrt{3}}\frac{m_{4}}{m_{2}},
\end{equation*}%
where 
\begin{equation*}
m_{n}=\int d^{2}q~q^{n}C\left( \mathbf{q}\right) .
\end{equation*}%
Using these equations we calculate $\beta \approx 2~\mathrm{nm}$ and $\beta
nh_{\mathrm{rms}}\approx 625$, whereas the value adopted by Masjedi and
Khonsari \cite{Khonsari} are $\beta = 0.02 \ \mathrm{mm}$ and $n\beta \sigma
=0.05$. Thus, the model parameters adopted in Ref. \cite{Khonsari} are
orders away from those really describing the true contact system; as a
result, the comparison reported in \cite{Khonsari} have to be reconsidered, so that further clarifying research is
needed on this aspect. In addition we note that a certain discrepancy can be
observed also under elasto-hydrodynamics conditions, when no
asperity-asperity or asperity-fluid interactions are relevant, reported on
in Fig. 1a of Ref. \cite{Khonsari}. This is likely due to a different
numerical inlet distance adopted by Masjedi and Khonsari, which might
sensibly affect the numerical results, as confirmed by the authors. Thus, it
would have been more helpful to provide numerical results for exactly the
same contact conditions.

\vskip0.5cm \textbf{3 Experimental setup} 
\label{setup}

\begin{figure}[tbp]
\begin{center}
\includegraphics[width=1.0\columnwidth]{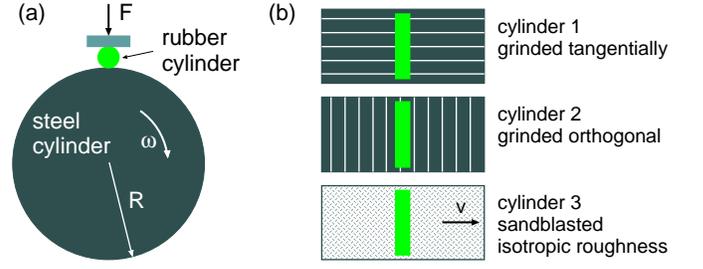}
\end{center}
\caption{(a) Schematic picture of the experimental friction tester. The
rubber cylinder is pushed with a dead weight towards the rotating steel
cylinder. (b) The steel cylinder has surface roughness prepared by grinding
the cylinder surface along the rotation direction (top), orthogonal to the
rotation direction (middle) or by sandblasting (bottom). The latter results
in surface roughness with isotropic statistical properties.}
\label{1}
\end{figure}

In order to experimentally investigate the lubricated line contact of a
generic hydraulic seal, a test rig has been designed and set up at the
Institute for Fluid Power Drives and Controls (IFAS). Here we summarize the
experimental setup, whose detailed description can be found in another
contribution \cite{conf.23}. A steel cylinder (radius $R=10\ \mathrm{cm}$)
is rotated through a two-stepped gearbox (with $\tau _{2}/\tau _{1}=0.025$
being the relative transmission ratio) by an electric motor (connected to a
frequency converter), whereas a strait segment of a nitrile butadiene rubber
(NBR) O-ring (length $40$ \textrm{mm}) is squeezed with a normal force $F$
against the steel surface (see Fig. \ref{1}(a)). The cross section of the
(undeformed) O-ring is circular, with the radius $r=2.5\ \mathrm{mm}$. The
relative sliding velocity can be varied in the range $2$ to $25~\mathrm{mm/s}
$ or $0.09$ to $1~\mathrm{m/s}$, depending on the selection of the gear
ratio. The adopted lubricant is a standard HLP 46 hydraulic oil, with
kinematic viscosity $\nu =137~\mathrm{mm%
{{}^2}%
/s}$ and oil density $\rho =875~\mathrm{kg/m}^{3}$ at ambient temperature.
Finally, both rotating cylinder and seal specimen are located in a
temperature-controlled bath chamber, which can be entirely flooded with a
lubricant, at environment pressure, and with temperature adjustable in the
range $10$ and $100~\mathrm{%
{{}^\circ}%
C}$.

The experimental results, to be shown in in Sec. 5, were performed at a
constant temperature of $20\pm 2^{\circ }\mathrm{C}$, whereas 5 different
velocities are adopted in the present study ($2.5~\mathrm{mm/s}$, $4.5~%
\mathrm{mm/s}$, $8~\mathrm{mm/s}$, $14~\mathrm{mm/s}$ and $25~\mathrm{mm/s}$%
), as well as three different loads ($31.1$, $93.3$, and $155.5~\mathrm{N}$,
corresponding to Hertz's pressures ($\tilde 1.2$, $2.1$, and $2.9~\mathrm{MPa}$)
are applied for each velocity.

\begin{figure}[tbp]
\begin{center}
\includegraphics[width=1.0\columnwidth]{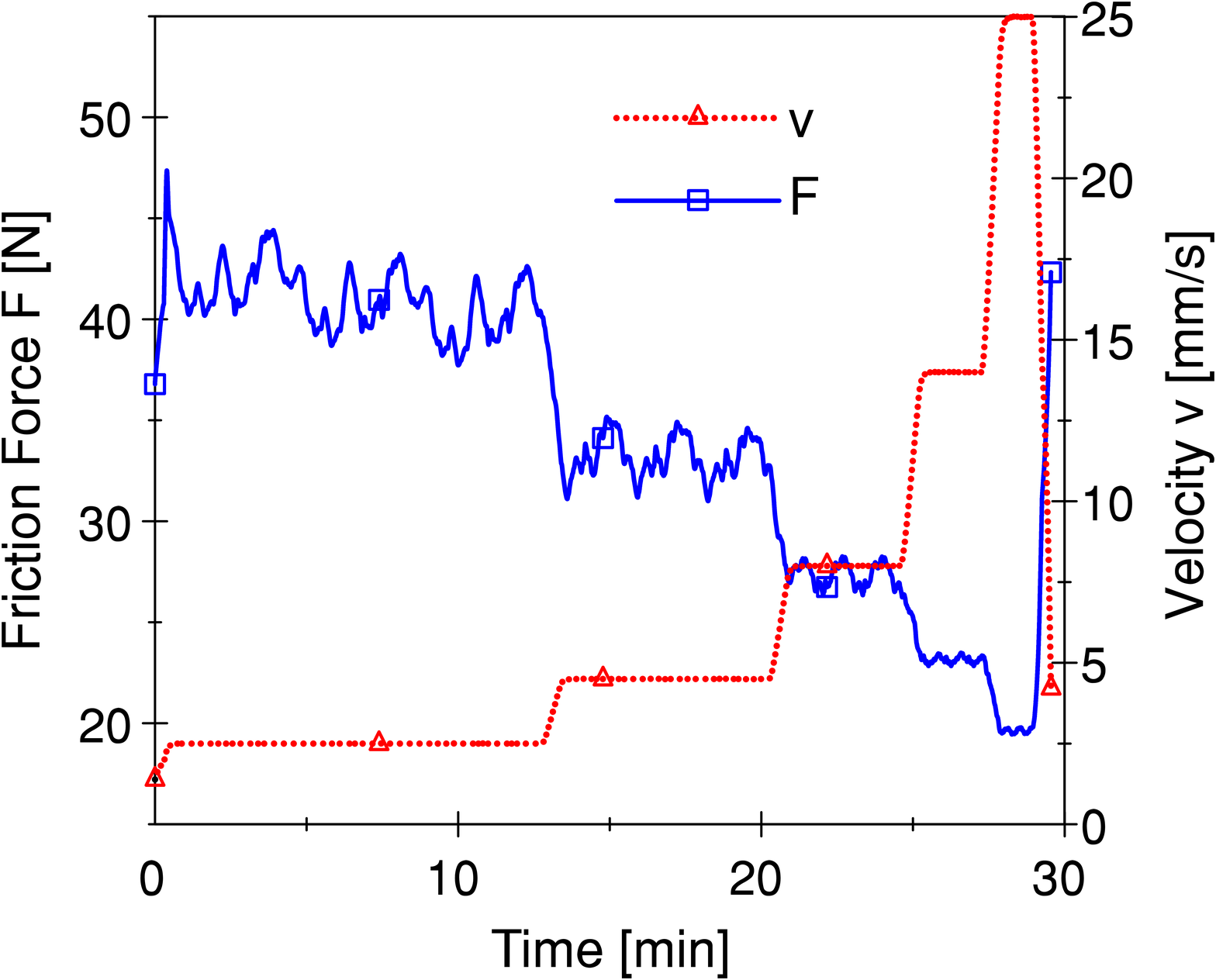}
\end{center}
\caption{Friction force and velocity as a function of time during a
typical experiment.}
\label{10}
\end{figure}
The test protocol, for each adopted rough surface (see in the following), is
as follows. Under constant load, the sliding velocity is kept constant and
friction measured during three cylinder revolutions before switching to the
next velocity, leading to a total measurement time of about $30~\mathrm{min}$%
. This procedure is repeated three times using the same contact pair in
order to detect possible running-in or wear effects of the seal. Afterward,
a new seal and new load is used. The same lubricant is used during all
experiments, however additional (repeated) experiments made at the end of
the experimental campaign have confirmed that no measurable aging or
contamination of the oil was occurring in the setup. An example chart
showing the temporal evolution of the sliding speed, and corresponding
measured friction force, is reported in Fig. \ref{10}.

\begin{table}[hbt]
\caption{Normal load and maximum contact pressure (according to Hertz
theory) during experiments.}
\label{tab:data}
\begin{center}
\begin{tabular}{|c||c|}
\hline
Normal force (N) & \ Maximum cont. pressure (MPa) \\ \hline\hline
31.1 & 1.2 \\ \hline
93.3 & 2.1 \\ \hline
155.5 & 2.9 \\ \hline
\end{tabular}%
\end{center}
\end{table}



\vskip0.5cm \textbf{4 Surface roughness power spectra} 
\label{Cq}

In all the experiments the contact consists of a steel cylinder with a rough
surface, whereas the rubber can be considered smooth. Three cylinders with
different surfaces were adopted. All the cylinder surfaces were first
sandblasted to provide a statistically similar initial roughness. Next, the
surfaces of two of the cylinders are grinded orthogonally (in the
following, cylinder 2) or longitudinally (cylinder 1) to the sliding
direction (perpendicular to the cylinder axis). A sketch of the surfaces is
shown in Fig. \ref{1}(b). 
The surface roughness was measured at different positions along the
azimuthal and meridian directions, through $20~\mathrm{mm}$ line-scans
(eight scans for each surface). For the sandblasted cylinder $h_{\mathrm{rms}%
}\approx 1.0~\mathrm{\mu m}$, and for the surfaces with anisotropic
roughness $h_{\mathrm{rms}} \approx 0.6~\mathrm{\mu m}$. The roughness is
higher then recommended for a standard hydraulic cylinder rod, but is well
suited for the investigation we provide in the following.

In the mean field theory it enters the two-dimensional (2D) power spectral density (PSD) $C(%
\mathbf{q})$ which can be calculated from the measured height profile $h(%
\mathbf{x})=h(x,y)$ using: 
\begin{equation*}
C\left( \mathbf{q}\right) =\left( 2\pi \right) ^{-2}\int d^{2}x\left\langle
h\left( 0\right) h\left( \mathbf{x}\right) \right\rangle e^{-i\mathbf{q}%
\cdot \mathbf{x}},\eqno{(12)}
\end{equation*}%
where the wave vector $q=(q_{x},q_{y})$, and where $h(\mathbf{x})$ is the
surface roughness height at the point $\mathbf{x}=(x,y)$, with $\left\langle
h\left( \mathbf{x}\right) \right\rangle =0$. However, (using a stylus
profiler) we have only measured the topography $h=h(x)$ along a line, and in
this case one can only calculate the one-dimensional power spectrum $C_{%
\mathrm{1D}}(q)$. For surfaces with isotropic roughness, $C_{\mathrm{2D}%
}\left( q\right) $ can be directly linked to line scans $C_{\mathrm{1D}}(q)$
using\cite{Carbone2009} 
\begin{equation*}
C_{\mathrm{2D}}\left( q\right) =-\int_{q}^{\infty }\frac{d k}{\pi} ~{\frac{C_{\mathrm{1D}%
}^{\prime }( k ) }{\sqrt{k^{2}-q^{2}}}}.
\end{equation*}

\begin{figure}[tbp]
\begin{center}
\includegraphics[width=1.0\columnwidth]{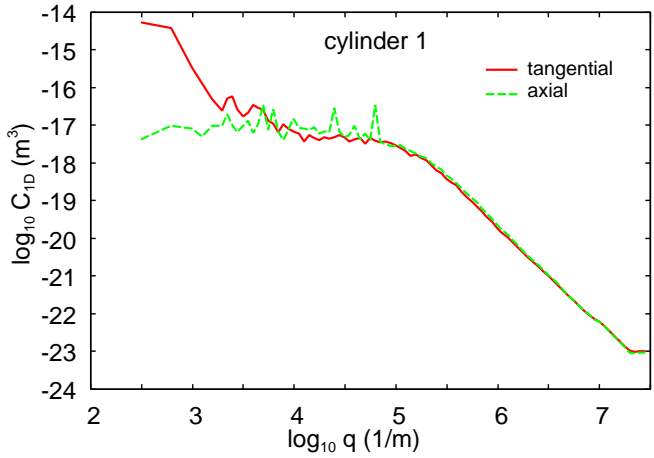}
\end{center}
\caption{1D surface roughness power spectra $C_{\mathrm{1D}}$ of cylinder
surface 1 as a function of the wavenumber q; axial direction (dashed line)
and the tangential direction (solid line).}
\label{7}
\end{figure}

\begin{figure}[tbp]
\begin{center}
\includegraphics[width=1.0\columnwidth]{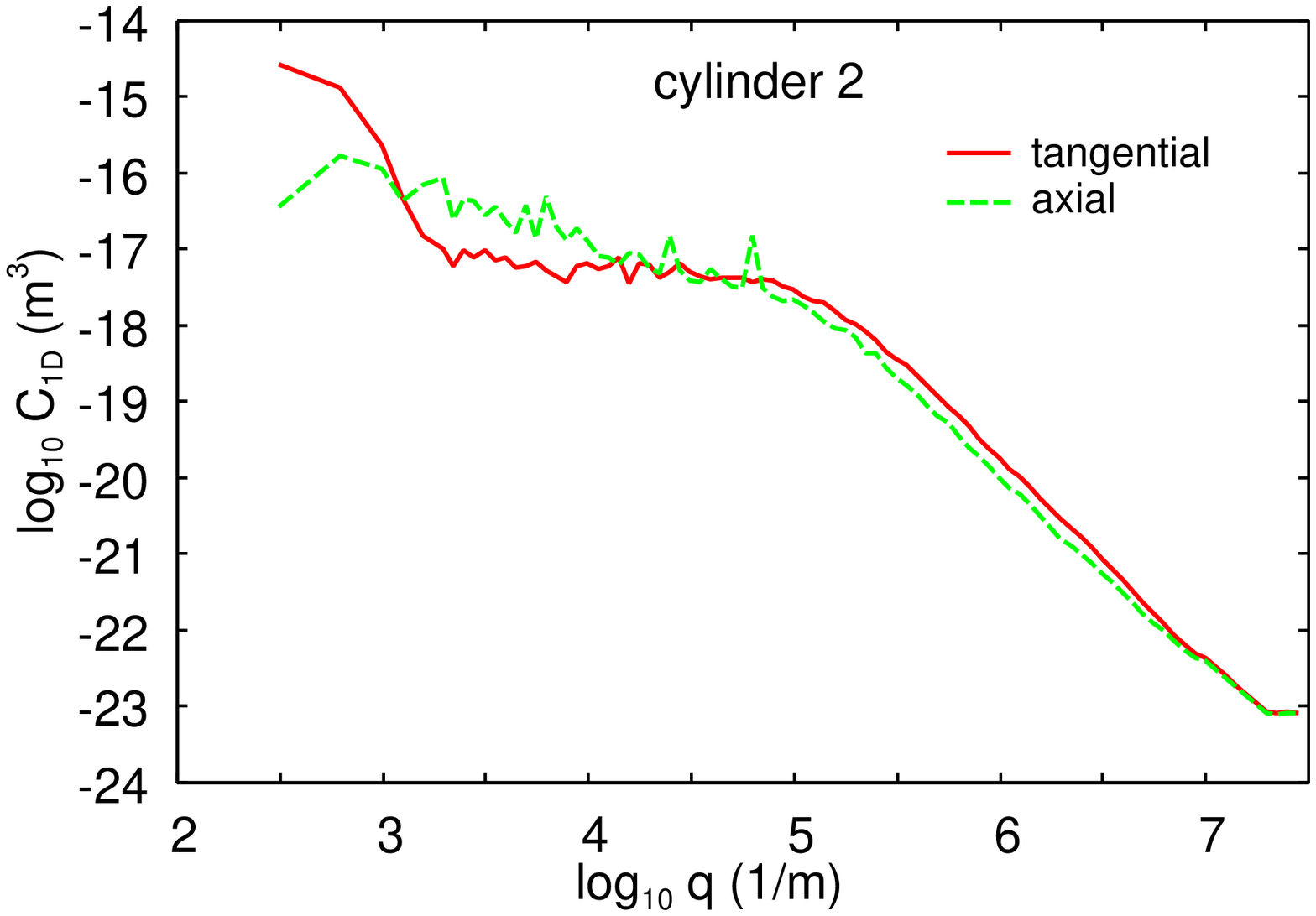}
\end{center}
\caption{1D surface roughness power spectra $C_{\mathrm{1D}}$ of cylinder
surface 2 as a function of the wavenumber q; axial direction (dashed line)
and the tangential direction (solid line).}
\label{8}
\end{figure}

\begin{figure}[tbp]
\begin{center}
\includegraphics[width=1.0\columnwidth]{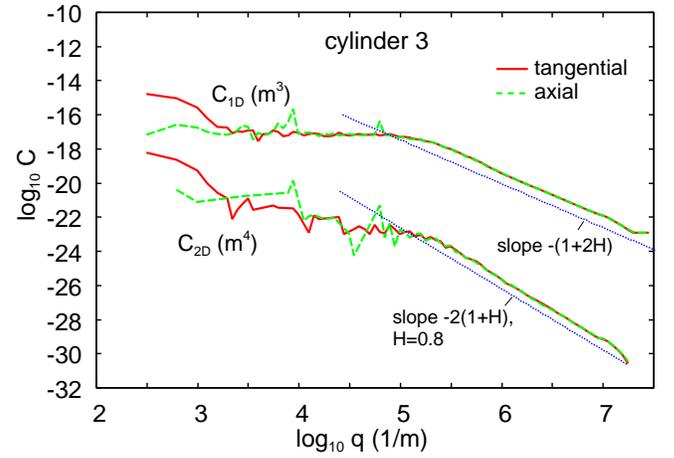}
\end{center}
\caption{1D and 2D surface roughness power spectra of cylinder surface 3 as
a function of the wavenumber q along the axial direction (dashed line) and
the tangential direction (solid line). The solid blue lines indicate the
slope of the corresponding self-affine region of the PSDs; they have been
drawn displaced from experimental curves only for the sake of readability.}
\label{9}
\end{figure}

In Fig. \ref{7}, \ref{8} and \ref{9} we show, respectively, the calculated
(averaged over 8 measurements repetitions) one dimensional surface roughness
power spectra $C_{\mathrm{1D}}$ as a function of the wavenumber on log-log
scale for cylinder 1 (longitudinal roughness), 2 (transverse roughness) and
3 (isotropic roughness). In the figures, the solid (red) and dashed (green)
lines are, respectively, from line-scan along and perpendicular to the
sliding direction. On the log-log scale, the PSDs are approximately linear
for $q\gtrapprox 10^{5}~\mathrm{m}^{-1}$, and in particular the surfaces
appear to be self affine fractal with the fractal dimension $D_{\mathrm{f}%
}=3-H\approx 2.2$, where $H$ denotes the Hurst exponent. Moreover, in Fig. %
\ref{9} (isotropic roughness) we also show the 2D surface roughness power
spectra, the latter calculated from the formula reported above. The solid
blue lines indicate the slope of the corresponding self-affine region of the
PSDs; they have been drawn displaced from experimental curves only for the
sake of readability.

\vskip0.5cm \textbf{5 Experimental results and discussion} \label%
{Experimental results}

\vskip0.2cm \textbf{5.1 Influence of squeezing load} 

\begin{figure}[tbp]
\begin{center}
\includegraphics[width=1.0\columnwidth]{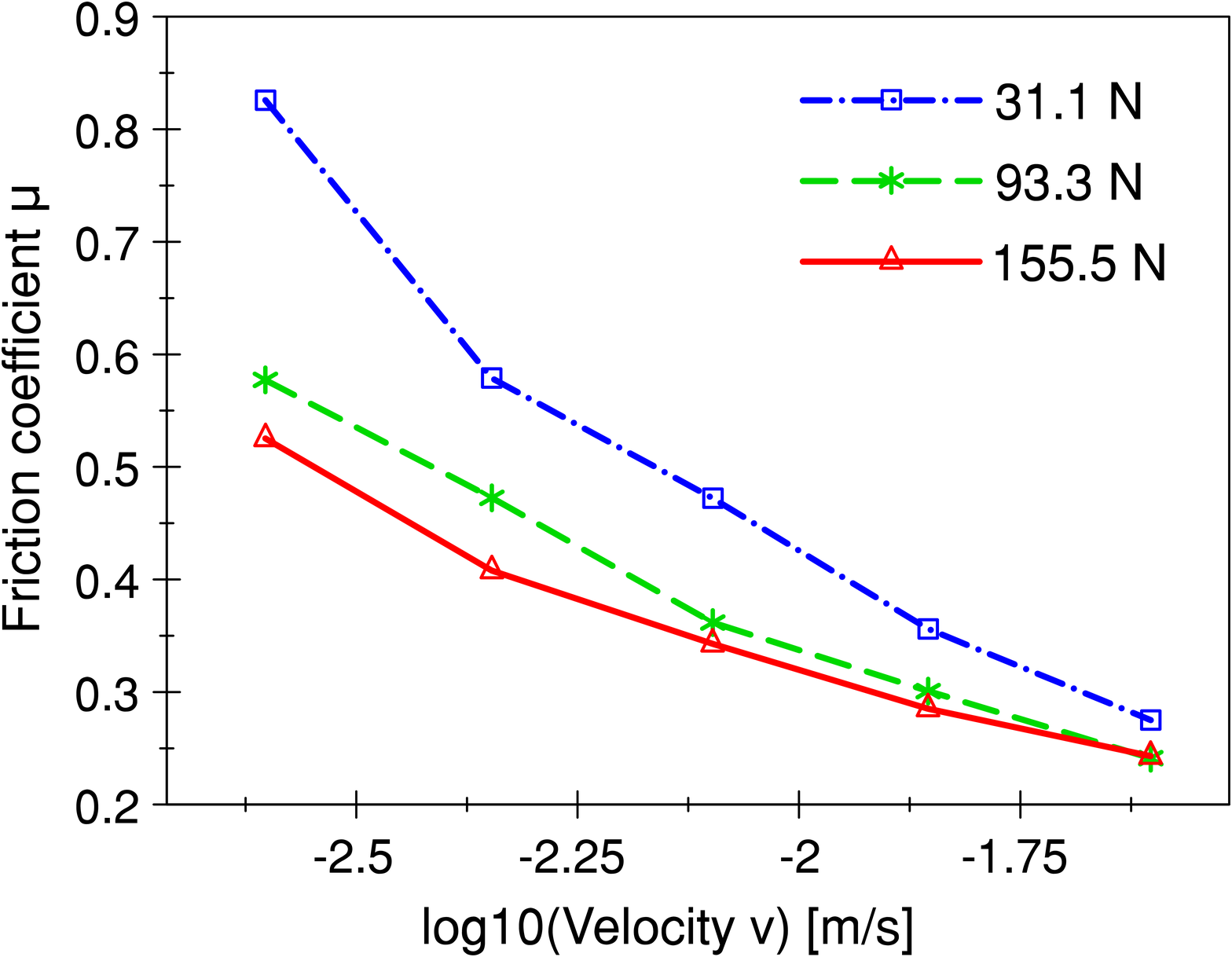}
\end{center}
\caption{Friction coefficient $\protect\mu$ as a function of the sliding
velocity $v_{0}$ for all three normal loads. For cylinder 2 (transverse
roughness).}
\label{11}
\end{figure}

In Fig. \ref{11} we show for cylinder 2 (transverse roughness) the friction
coefficient ${\mu}$ as a function of the sliding velocity $v_{0}$ for the
three different loads. The plotted friction coefficient is the arithmetic
mean of all corresponding measurements (error bars are not included in Fig. %
\ref{11} for the sake of readability, but they will be included in the
results following in this section). 
Notice that the friction coefficient $\mu$ decreases as the velocity
increases. Considering the adopted contact geometry and distribution of
roughness and compliance on the solids, this behavior can be simply ascribed
to the occurrence of the mixed lubrication regime in the contact.
Furthermore, the friction coefficient decreases as the normal load
increases. This too can be justified by observing that, in the mixed regime,
part of the load is supported by the fluid-asperity interactions. Thus by,
e.g., linearly increasing the load, the asperity contact area does not
increase linearly with the load, and the tangential force, which is given by
a a contact area term and a fluid shear term, increases less than linearly
with increasing load, resulting in the lower friction coefficient (see Sec. 6).

\vskip0.2cm \textbf{5.2 Influence of the surface topography} 

\begin{figure}[tbp]
\begin{center}
\includegraphics[width=1.0\columnwidth]{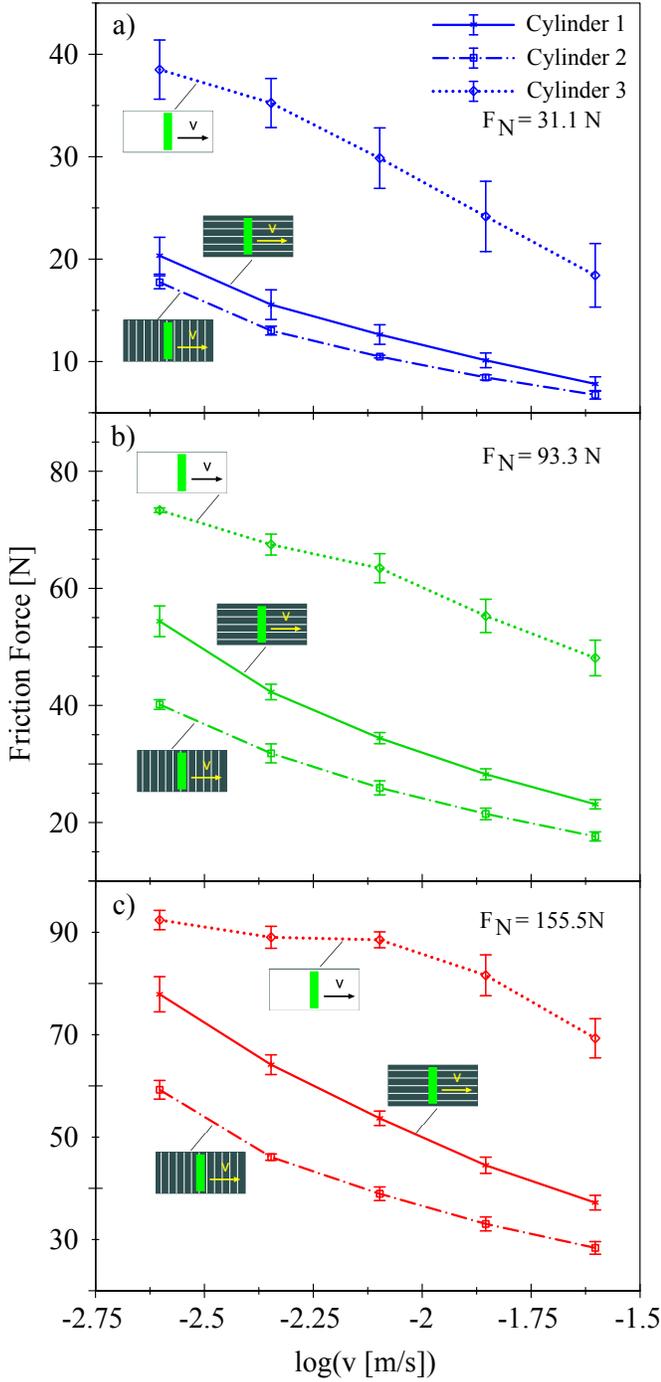}
\end{center}
\caption{Friction forces as a function of the velocity $v$ (log scale), for
all the test cylinders. Normal load (a) $31.1N$, (b) $93.3N$ and (c) $155.5N$%
. }
\label{friction.force.all.eps}
\end{figure}

Here we present the measured friction forces as a function of the sliding
speed, for all three cylinders, at a normal load of $31.1~\mathrm{N}$ [Fig. %
\ref{friction.force.all.eps}(a)], $93.3~\mathrm{N}$ [Fig. \ref%
{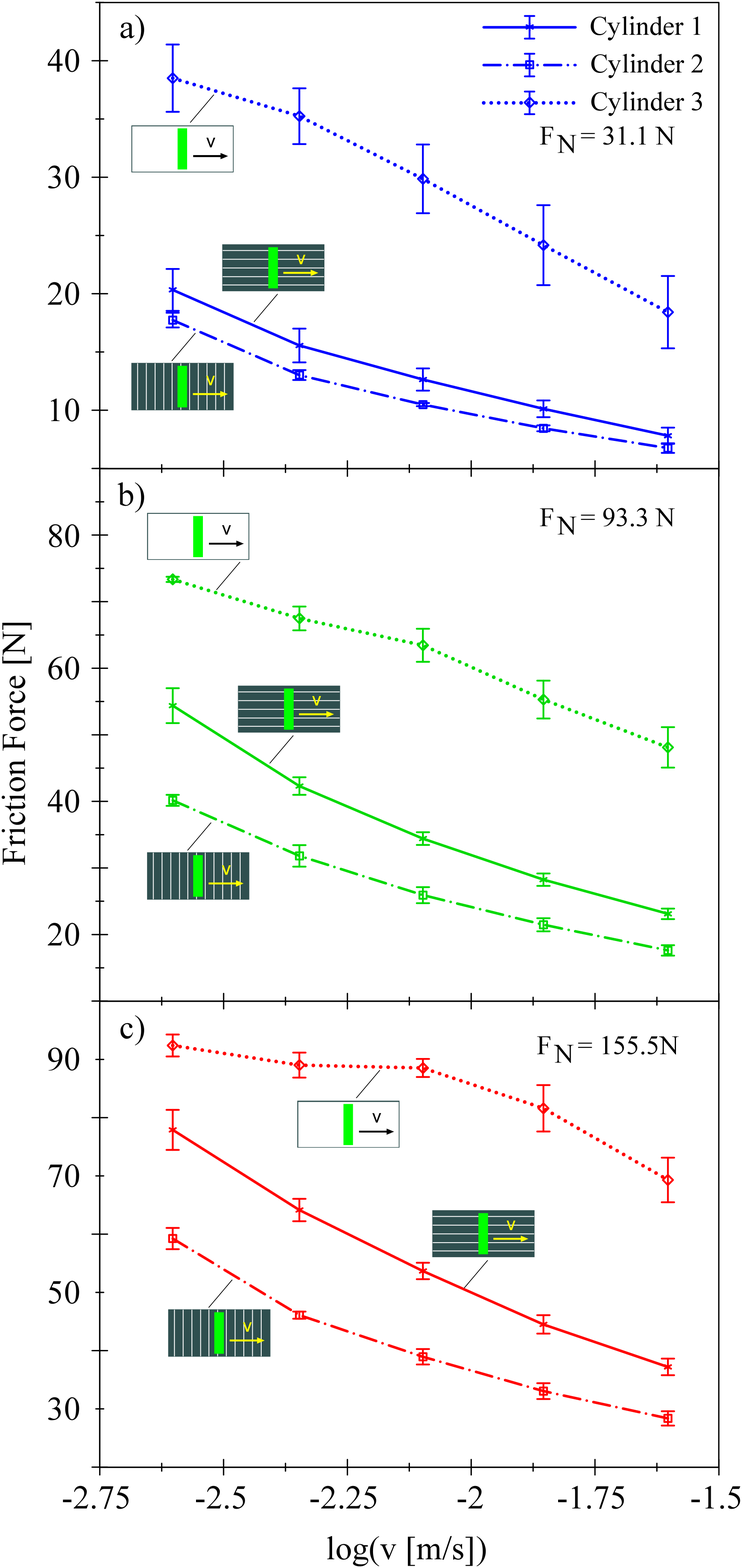}(b)] and $155.5~\mathrm{N}$ [Fig. \ref%
{friction.force.all.eps}(c)]. 
The friction forces for cylinder 3 (isotropic roughness) are significantly
higher, compared to the forces for cylinder 1 (longitudinal roughness) and 2
(transverse roughness). This is expected by the larger roughness,
which makes the asperity-asperity interactions more important up to larger
sliding velocities than for the lower roughness surfaces. Thus, at a same
sliding speed, a larger solid contact area, and a larger solid contact
friction, occurs for the rougher cylinder. We note that this is only true in
the high velocity part of the mixed lubrication region, since in the boundary
lubrication region the larger roughness result in smaller contact area and
hence a smaller friction, at least if one can neglect the viscoelastic
contribution to the friction 
(see Fig. \ref{1logv.2muTOT.eps} which illustrate this point).

It is interesting to compare the friction forces occurring for the contact case
1 and 2. In particular, the dissipation occurring for the cylinder with
longitudinal roughness is always larger than for the transverse roughness.
Considering that the rough surfaces show similar high (same Hurst exponent)
and low (same root mean square roughness) frequency content, this must be
due to the anisotropy of the surface. Indeed, as pointed out before,
the grinding wear tracks along the sliding direction facilitate the
removal of the fluid from the nominal contact region, and extend the
boundary lubrication region to higher sliding speeds. In most cases this
results in an increase in the friction force compared to the case of
isotropic roughness with the same angular averaged power spectrum. In a
similar way, when the grinding direction is orthogonal to the sliding
direction, the fluid removal is reduced, resulting (in most cases) to lower
friction than for isotropic roughness. This is also consistent with
the results in Fig. \ref{1logv.2muTOT.eps} which in the high velocity
part of the mixed lubrication part of the Stribeck curve show higher friction
when sliding along the grinding direction.

The observation that the friction force
of cylinder 1 is higher than the friction force of
cylinder 2 applies to all investigated velocities and to all tested
normal loads. Especially for the normal loads $93.3~\mathrm{N} $ and 
$155.5~\mathrm{N}$ the difference is significant. As the grinding direction
is the only difference, the measured effect has to be caused by the
anisotropic surfaces. In particular, the friction curves of cylinder 3, in
the log-linear representation, show a negative curvature with respect to the
(positive) curvature shown by the other surfaces.

At very low sliding speeds (in the boundary lubrication region) it is likely
that the friction force is biggest when the grinding direction is orthogonal
to the sliding direction. This is the case because fluid hydrodynamic
effects are not important at very low sliding speed. At the same time there
may be a viscoelastic contribution to the friction which will be largest
when sliding orthogonal to the grinding direction 
(see discussion around Fig. \ref{1logv.2muTOT.eps}); with the present
experimental set-up we where not able to study the friction in the boundary
lubrication velocity range.

To summarize, the friction asymmetry exhibited in Fig. \ref%
{friction.force.all.eps} is related to the anisotropy of the adopted surface
roughness, where surface 1 has a main scratch direction which is aligned
with the sliding direction. This configuration has been shown in the
literature to provide the smallest fluid-asperity interaction contribution
([12], [13], [15] and [16]): roughly speaking, the fluid is allowed to
escape along the roughness channels (through the scratch valleys) without
generate a strong fluid overpressure. Thus, the total fluid pressure is
almost negligible, resulting in a larger amount of asperity-asperity
interactions. This finally results in a Stribeck curve which occurs closer
to the boundary-mixed regime, the latter typically characterized by a
negative curvature. The opposite holds for the surfaces 2 and 3 where,
instead, the fluid-asperity interaction is higher, allowing to displace the
Stribeck curve toward the mixed-hydrodynamic regime, where the curvature is
typically positive. This is in perfect agreement with the results also shown
in Fig. \ref{19}.

\vskip0.2cm \textbf{5.3 Discussion}

The experiments show a distinct influence of the surface topography,
especially when comparing the two anisotropic surfaces 1 and 2. Two effects
do contribute to the friction process.

i) The fluid-asperity interactions occurring when the sliding direction is
perpendicular to the grooves main direction (so called transverse roughness)
is typically stronger than for the aligned grooves (longitudinal roughness),
when the groove representative size is smaller than the smallest nominal
contact length. This is well known in the literature of mixed lubrication
for soft contacts [11, 12] and of hydrodynamic lubrication for textured
surfaces [13, 14]. Thus, a larger asperity-induced fluid pressure is
provided for the transverse roughness, which in turn determines an increased
average interfacial separation (resulting from the decrease of the
solid-solid contact pressure) and, therefore, a reduced dry contact area. However, in term of the resulting
fluid viscous dissipation, this stronger fluid-asperity interaction for the
transverse roughness can cause both an increase or decrease of the total
friction, depending on the range of sliding velocity at which this
interaction dominates. Thus, in the present case the stronger fluid-asperity
interaction (for the transverse roughness range) only determines a reduction
of the dry-contact friction (due to the decrease of the contact area),
whereas the fluid-viscous dissipation occurring at the asperity scale
(micro-EHL) seems not to dominate this mixed lubrication contact
configuration (as a counter example, see Ref. [15]).

ii) The reduced dry (or boundary lubricated) contact area suggested in i)
determines, consequently, a reduction of the friction contributions coming
from the intimate (or boundary mediated) asperity-asperity interaction. In
particular, the solid-contact friction can originate, in the present case,
by both the adhesive contribution (proportional to the dry true contact
area) and by the micro-rolling contribution (a dissipation originating by
the pulsating deformation on the rubber seal by the sliding asperities)
[16]. For small true contact areas (of interest for the seal application),
and depending on the sliding speed, the hysteretic contribution to friction
is slightly dependent on the contact area. Thus, the decrease of dry contact
obtained for the transverse roughness does induce a decrease of dry
friction, resulting in the overall friction decrease with respect to the
longitudinal roughness.

\vskip 0.5cm \textbf{6 Theory analysis}

In this section we will analyze the experimental results presented above.
The friction coefficient as a function of sliding speed for the lubricated
contact, the so called Stribeck curve, will be calculated. We will discuss
in detail how the surface roughness influences the Stribeck curve, and in
particular under which conditions it will shift the Stribeck curve to lower
sliding speeds as this reduces the friction and the wear of the rubber seal.
As input for the calculation, the surface roughness power spectra of the
surfaces involved (Figures \ref{7}, \ref{8} and \ref{9}), and the effective
elastic modulus of the solids are needed. In the study below the surface
roughness on the rubber surface will be neglected, and only the roughness of
the steel cylinders is considered.

Due to the incomplete experimental information about the surface roughness
anisotropy, and the absence of information about the frictional properties
of the dry rubber-countersurface contacts, in particular how the frictional
shear stress in the area of real contact depends on the sliding speed, the
present study is only of semi-quantitative nature. Nevertheless, the origin
of the observed dependency of the friction coefficient on the load and the
influence of the surface roughness anisotropy on the friction will be
explained. The anisotropy of surface roughness can be characterized by the
Tripp number. Roughly speaking, the Tripp number
is the ratio $\gamma =\xi _{x}$/$\xi _{y}$ between the axis of the elliptic
cross-section (in the $xy$-plane) of an (average) asperity in real $(x,y)$
space. In fluid
dynamics, the Tripp number is the quantity which naturally appears in the
fluid flow equations. The Tripp number can be easily obtained from the 2D
surface roughness power spectrum $C(q_{x},q_{y})$, see Ref. [17, 11]. Thus,
to determine the Tripp number accurately, topography measurements over
surface areas rather than 1D line-scans are necessary. It is clear, however,
that in the present case, because of the preparation method, surface 3 has
the Tripp number $\gamma =1$, while surface 2 has $\gamma <1$. Since surface
1 is ground orthogonal to the cylinder axis, i.e., along the sliding
direction, it must have $\gamma >1$ but Fig. \ref{7} shows it must be very
close to one as nearly no difference occurs in the 1D power spectra along
the two directions.

For a complete and accurate analysis one need in general the surface
roughness power spectra for all wavenumbers, i.e., down to atomic distances
corresponding to wavenumbers $q\approx 10^{10}~\mathrm{m}^{-1}$, or
wavelength of order $\mathrm{nm}$, which requires studying the surface
roughness with (near) atomic resolution, using, e.g., Atomic Force
Microscopy (AFM). However, it will be shown below that the velocity and load
dependency of the friction force in the present case is due mainly to the
influence of the longest wavelength roughness components on the fluid
dynamics. The large wavenumber cut-off $q_{1}$ is chosen so that the surface
root mean square slope equal 1.3, when including all the roughness with
wavenumber $q<q_{1}$. The results presented below are not sensitive to this
assumption.

For the mixed-elastohydrodynamic calculations presented below the elastic
properties of the solids are needed. The steel cylinders will be treated as
rigid. Rubber is a viscoelastic material which can be characterized by a
viscoelastic modulus $E\left( \omega \right) $ which depends on the
frequency $\omega $ with which it is deformed. Note that $E\left( \omega
\right) $ is a complex quantity, where the imaginary part is associated with
energy dissipation, resulting from the friction force which acts between the
rubber chain molecules when they slip relative to each other. For lubricated
sliding friction the viscoelasticity of the rubber will also contribute to
the friction in the area of contact. However, in the present study this will
not be considered in detail but it will just be assumed that a constant
frictional shear stress $\tau_1$ acts in the area of real contact (see also
discussion below). Now, when the rubber cylinder slides on the steel counter
surface there will be two types of deformations. First, there will be a
macroscopic Hertz's-like deformation of the rubber cylinder. This
deformation is constant in time so it is determined by the low frequency
modulus of the rubber which was measured (including non-linearity effects)
to be about $E\approx 9~\mathrm{MPa}$. However, the true area of contact is
determined by the interaction between the asperities on the steel surface
and the rubber. During sliding this includes high-frequency deformations
involving a band of deformation frequencies $\omega \approx qv$, where $%
q=2\pi /\lambda $ is the wavenumber (and $\lambda $ the wavelength) of a
surface roughness component, and $v$ the sliding speed. This will result in
a stiffening of the rubber with increasing sliding speed, and to a contact
area $A(v)$ which, even for dry contact, decreases with increasing sliding
speed. In this study this effect will not be included but it is assumed that
in the boundary lubrication region $A\left( v\right) \tau \left( v\right) $,
i.e. the adhesive contribution to friction, is a constant. Thus, the only
velocity (and load) dependency in the figures presented below is due the
fluid dynamics. In reality, this may not hold accurately, but it is likely
that the strong velocity dependence observed in a relative narrow (mixed
lubrication) velocity region, is due mainly to fluid dynamics effects.

\begin{figure}[tbp]
\begin{center}
\includegraphics[width=1.0\columnwidth]{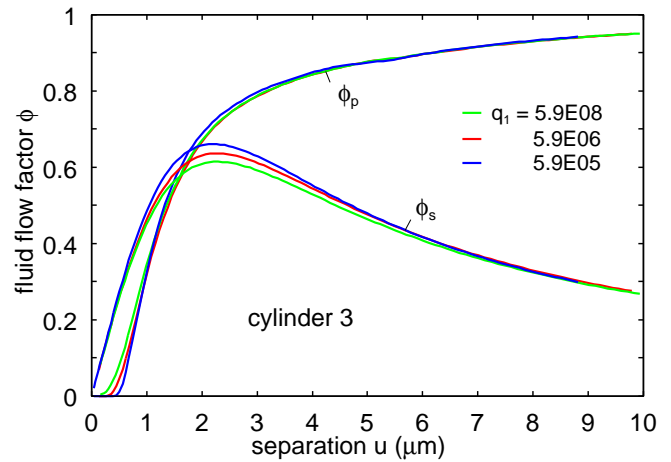}
\end{center}
\caption{The fluid pressure flow factor $\protect\phi_P$ and shear stress
flow factor $\protect\phi_s$ as a function of the surface separation using
the surface roughness power spectrum of cylinder 3 with three different
large wavenumber cut-off: $q_1 = 5.9~10^{8}$, $q_1 = 5.9~10^{6}$ and $q_1 =
5.9~10^{8}$ $m^{-1}$.}
\label{18}
\end{figure}

\begin{figure}[tbp]
\begin{center}
\includegraphics[width=1.0\columnwidth]{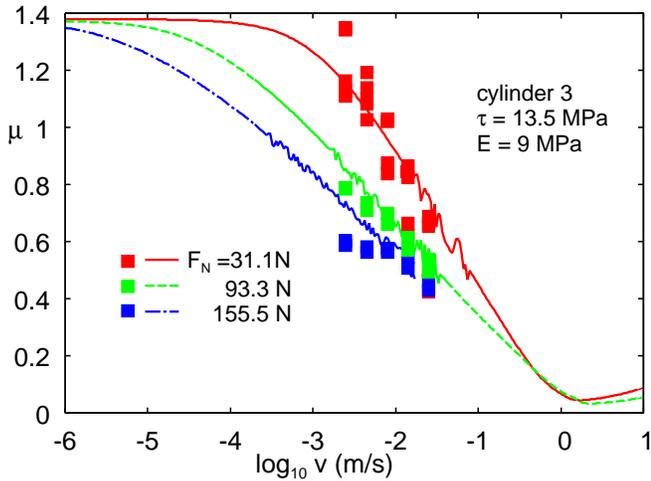}
\end{center}
\caption{The calculated (solid lines) and measured friction coefficients as
a function of sliding speed for three different loads, $F_{\mathrm{N}}$ =
31.1, 93.3 and 155.5 N.}
\label{19}
\end{figure}

We now consider the sliding of an elastic cylinder on the hard, lubricated
counter surface with random roughness. First the friction for cylinder 3 is
considered, which has surface roughness with isotropic statistical
properties. Fig. \ref{18} shows the fluid pressure and shear stress flow
factors, $\phi _{\mathrm{p}}$ and $\phi _{\mathrm{s}}$, as a function of the
surface separation $\bar{u}$, as obtained using the theory of Ref. [14].
Note that for isotropic roughness $\phi _{\mathrm{p}}<1$, and since the
pressure flow factor appears together with the viscosity in the combination $%
\eta _{\mathrm{eff}}=\eta /\phi _{\mathrm{p}}>\eta $, it follows that the
fluid pressure flow factor will shift the mixed lubrication region towards
lower sliding speed, as well as increase the strength of the generic
fluid-asperity interaction given the larger effective viscosity. When the
surface roughness occurs on the moving substrate surface, as in the present
case, the fluid shear stress flow factor $\phi _{\mathrm{s}}$ is positive,
and according to Eq. (2) this will result in a shift of the Stribeck curve
towards lower sliding speed. The physical reason for this is as follows:
consider the rough substrate surface moving against the smooth elastic
cylinder. The fluid carried in the valleys of the moving rough surface helps
to transport fluid from the leading edge into the gap, and hence support
fluid lubrication, and shift the Stribeck curve towards lower sliding speed.

A very important observation for what follows is that unless the contact
pressure is very high, the flow factors are mainly determined by the most
long-wavelength part of the surface roughness. This is illustrated in Fig. %
\ref{18} where the flow factors for three different large wavenumber cut-off 
$q_{1}$ are shown. Note the relative small difference between the three
cases in spite of the large variation (by three decades in length scale) in
the cut-off $q_{1}$. The reason for this is that the long-wavelength (and
large amplitude) surface roughness components generate the biggest fluid
flow channels at the interface, and will hence influence the fluid flow
dynamics at the interface much stronger than the smaller channels arising
from the shorter surface roughness components. At very high contact
pressures this is no longer the case, but for the present situation it holds
to a very good approximation.

To obtain the Stribeck curve for cylinder 3, information about the frictional
shear stress acting in the area of real contact is needed. For rubber
sliding on dry hard rough surfaces the friction coefficient is usually
nearly independent of the nominal contact pressure [7]. In particular, the
frictional shear stress $\tau $ acting in the area of real contact is
independent of the local contact pressure in the pressure range relevant in
most applications involving rubber materials. For lubricated surfaces it was
found above that the friction coefficient depends strongly on the load. In
the hydrodynamic region for smooth surfaces the Stribeck curve depends on $%
\eta v/F_{\mathrm{n}}$ but this scaling is not valid in the mixed
lubrication region. In Fig. \ref{19} the relation between the friction
coefficient and the sliding speed for cylinder 3 and for the three different
loads used in the experiments reported on above are shown. In the
calculations was assumed that a (velocity independent) frictional shear
stress $\tau_1 =13.5~\mathrm{MPa}$ acts in the area of real contact.

\begin{figure}[tbp]
\begin{center}
\includegraphics[width=1.0\columnwidth]{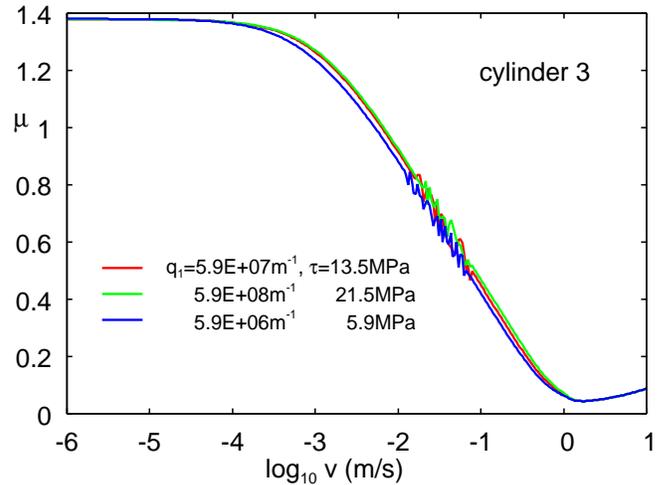}
\end{center}
\caption{The calculated friction coefficients as a function of sliding speed
for the load $F_{\mathrm{N}}$ = 31.1 N for several different large
wavenumber cut-off q1.}
\label{20}
\end{figure}

Note that the friction coefficient depends on the load, in relative good
agreement with the experimental data (square symbols). Due
to a numerical instability, for the highest load the calculated results
terminates before reaching the highest sliding speed. As pointed out above,
unless the contact pressure is very large (resulting in very small
interfacial separation), the fluid flow factors depend mainly on the longest
wavelength roughness components. However, the area of real contact depends
on all the roughness wavelength components and decreases continuously as the
large wavenumber cut-off $q_{1}$ increases. In the boundary lubrication
region all the load is carried by the area of real contact $A$ and in this
case in our model the friction force $F_{\mathrm{f}}=A\tau $ (note: in
reality, in the boundary lubrication velocity range there will also be a
viscoelastic contribution to the friction from the time-dependent
deformations of the rubber surface by the asperities of the countersurface).
In Fig. \ref{20} the Stribeck curves for cylinder 3 are shown, using three
different large wavenumber cut-off $q_{1}$ but adjusting the frictional
shear stress so the friction force in the boundary lubrication region is
(nearly) the same in all cases. Note that in this case the Stribeck curves
are nearly identical. This is the case only because the fluid flow factors
are mainly determined by the most long-wavelength part of the surface
roughness spectra, which is the same in all cases.

\begin{figure}[tbp]
\begin{center}
\includegraphics[width=1.0\columnwidth]{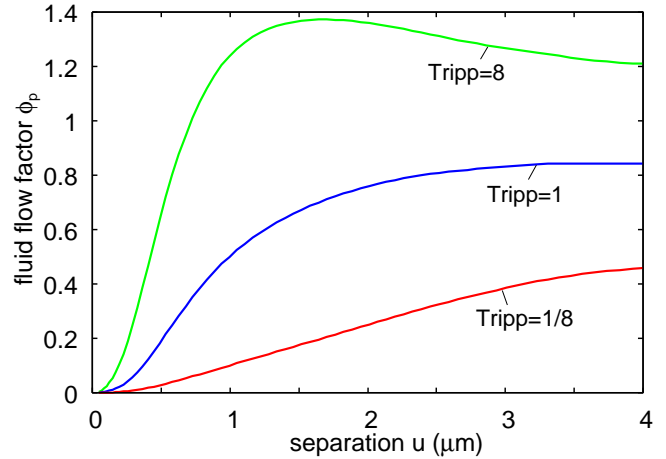}
\end{center}
\caption{The fluid pressure flow factor $\protect\phi_P$ as a function of
the average separation u for cylinder 2 assuming the Tripp asymmetry number $%
\protect\gamma$ = 1, 8 and 1/8. }
\label{21}
\end{figure}

\begin{figure}[tbp]
\begin{center}
\includegraphics[width=1.0\columnwidth]{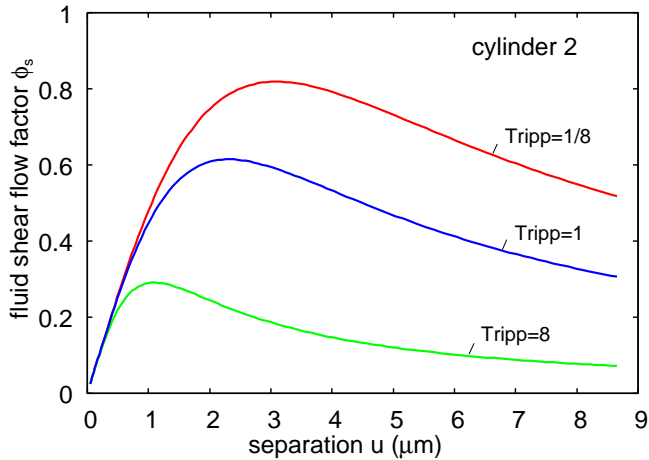}
\end{center}
\caption{The fluid shear stress flow factor $\protect\phi_S$ as a function
of the average separation u for cylinder 2 assuming the Tripp asymmetry
number $\protect\gamma$ = 1, 8 and 1/8. }
\label{22}
\end{figure}

We now consider the rubber friction for cylinder 2. For this case the fluid
flow factors for three different cases are calculated, namely isotropic
roughness (corresponding to the Tripp number $\gamma =1$), and for
anisotropic roughness with the grinding groves along the sliding direction
(corresponding to the Tripp number $\gamma =8$), and with the grinding
direction orthogonal to the sliding direction (corresponding to the Tripp
number $\gamma =1/8$) as in the actual system. Fig. \ref{21} and \ref{22}
show the fluid flow and shear stress flow factors for all three cases,
respectively.

\begin{figure}[tbp]
\begin{center}
\includegraphics[width=1.0\columnwidth]{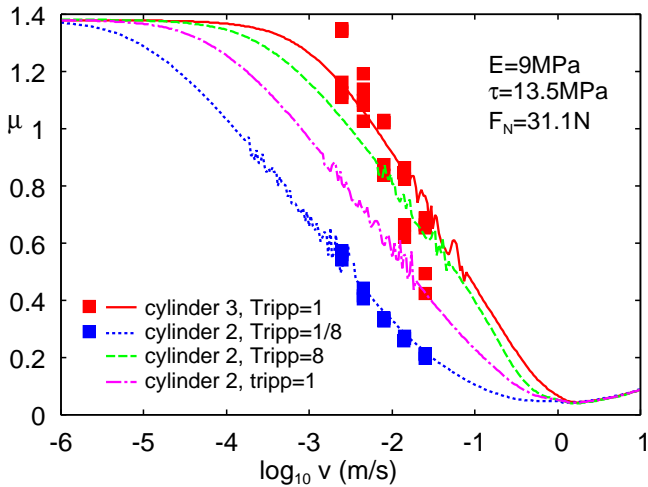}
\end{center}
\caption{The calculated (solid lines) and measured friction coefficients as
a function of sliding speed for the load $F_{\mathrm{N}}$ = 31.1 N.}
\label{23}
\end{figure}

Note that for $\gamma =8$, for the surface separation $\bar{u}>1~\mathrm{\mu
m}$ the fluid flow factor $\phi _{\mathrm{p}}>1$, which would shift the
Stribeck curve towards higher sliding speeds. However, in the studied
velocity range the interfacial separation is smaller than $0.6~\mathrm{\mu m}
$, in which case the pressure flow factor is smaller than 1 (see Fig. \ref%
{21}). As a result the fluid pressure flow factor will shift the Stribeck
curve to lower sliding speeds. Similarly, the shear stress flow factor
shifts the Stribeck curve towards lower sliding speeds. For the surface with
the grinding direction orthogonal to the sliding direction (which is the
actual case for cylinder 2), $\gamma =1/8$, $\phi _{\mathrm{p}}<1$ for all
interfacial separations, and both the pressure and shear stress flow factors
shift the Stribeck curve towards lower sliding speed, see Fig. \ref{23}.
Fig. \ref{23} also show the calculated Stribeck curve for the cylinder 3 is
shown (experimental data from Fig. \ref{19}).

Note that in the studied velocity interval the friction for cylinder 3 is
larger than for cylinder 2, which is due to the larger roughness on cylinder
3, and also due to the difference in the surface roughness anisotropy, both
of which result in faster fluid squeeze-out for cylinder 3.

\vskip 0.5cm \textbf{7 Summary and conclusions} 

In this paper we have presented a combined experimental-theoretical investigation
of the influence of anisotropic surfaces on the friction behavior of a
hard-soft contacts. We have discussed in detail how the fluid flow and
friction shear stress factors depend on the surface roughness. A test rig
was described, which is designed to investigate a soft, lubricated line
contact in detail. An O-ring cord is brought into contact with the lateral
surface of an uniformly rotating steel cylinder. Three cylinders with different
surfaces are used: One sandblasted isotropic surface and two anisotropic
surfaces, grooved orthogonal or longitudinal to the direction of motion. A
distinct influence of anisotropic surfaces is detected. The friction force
decreases when the surface is grooved perpendicular to the direction of
motion. Finally, the experimental results were
compared with calculations.
These experimental results clearly show that the anisotropy of
surfaces in engineering applications has to be considered when modeling
lubricated friction of hydraulic seals. 

\begin{acknowledgments}
The research work was performed within a Reinhart-Koselleck project funded
by the Deutsche Forschungsgemeinschaft (DFG).
We would like to thank DFG for the project support
under the reference German Research Foundation DFG-Grant: MU 1225/36-1.
The research work was also supported by the DFG-grant: PE 807/10-1.
Finally, MS also acknowledges COST Action MP1303 for
grant STSM-MP1303-171016-080763.
\end{acknowledgments}

\appendix

\begin{widetext}

\section{}
Here we show how to calculate the rolling friction (both micro- and
macro-contributions) coming from the solid contact. We first observe that
the following contact relation applies%
\begin{equation*}
p_{\mathrm{cont}}\left( \mathbf{x}\right) \nabla u\left( \mathbf{x}\right) =0,
\end{equation*}%
leading to the equality%
\begin{equation*}
p_{\mathrm{cont}}\left( \mathbf{x}\right) \nabla u^{\mathrm{top}}\left( 
\mathbf{x}\right) =p_{\mathrm{cont}}\left( \mathbf{x}\right) \nabla u^{%
\mathrm{down}}\left( \mathbf{x}\right) ,
\end{equation*}%
where $u^{\mathrm{top}}\left( \mathbf{x}\right) $ and $u^{\mathrm{down}%
}\left( \mathbf{x}\right) $ correspond to the top- and down-body surface,
respectively. $p_{\mathrm{cont}}\left( \mathbf{x}\right) \nabla u^{\mathrm{%
down}}\left( \mathbf{x}\right) $ provides on the bottom solid (same but with
minus sign on the upper solid) the tangential component of the
locally-applied solid contact pressure, which contributes to the total
measured sliding friction. We are interested in the ensemble average $%
\left\langle p_{\mathrm{cont}}\left( \mathbf{x}\right) \nabla u^{\mathrm{down%
}}\left( \mathbf{x}\right) \right\rangle $ which reads%
\begin{equation*}
\left\langle p_{\mathrm{cont}}\left( \mathbf{x}\right) \nabla u^{\mathrm{down%
}}\left( \mathbf{x}\right) \right\rangle =\frac{A_{\mathrm{c}}}{A_{0}}%
\left\langle p_{\mathrm{cont}}\left( \mathbf{x}\right) \nabla u^{\mathrm{down%
}}\left( \mathbf{x}\right) \right\rangle _{A_{\mathrm{c}}},
\end{equation*}%
where $\left\langle {}\right\rangle _{A_{\mathrm{c}}}$ is the ensemble
average provided on the solid contact domain. We write%
\begin{equation*}
\left\langle p_{\mathrm{cont}}\left( \mathbf{x}\right) \nabla u^{\mathrm{down%
}}\left( \mathbf{x}\right) \right\rangle _{A_{\mathrm{c}}}=\frac{\bar{p}_{%
\mathrm{cont}}\left( \mathbf{x}\right) }{A_{\mathrm{c}}/A_{0}}\left\langle
\nabla u^{\mathrm{down}}\left( \mathbf{x}\right) \right\rangle _{A_{\mathrm{c%
}}}+\left\langle p_{1,\mathrm{cont}}\left( \mathbf{x}\right) \nabla u_{1}^{%
\mathrm{down}}\left( \mathbf{x}\right) \right\rangle _{A_{\mathrm{c}}},
\end{equation*}%
$\frac{A_{\mathrm{c}}}{A_{0}}\left\langle p_{1,\mathrm{cont}}\left( \mathbf{x%
}\right) \nabla u_{1}^{\mathrm{down}}\left( \mathbf{x}\right) \right\rangle
_{A_{\mathrm{c}}}=\left\langle p_{1,\mathrm{cont}}\left( \mathbf{x}\right)
\nabla u_{1}^{\mathrm{down}}\left( \mathbf{x}\right) \right\rangle $ is the
micro-rolling friction (to be calculated as shown in Ref. XX), whereas $\bar{%
p}_{\mathrm{cont}}\left( \mathbf{x}\right) \left\langle \nabla u^{\mathrm{%
down}}\left( \mathbf{x}\right) \right\rangle _{A_{\mathrm{c}}}$ is the
macro-rolling friction. In (10a) and (10b) the term $\left\langle p_{1,%
\mathrm{cont}}\left( \mathbf{x}\right) \nabla u_{1}^{\mathrm{down}}\left( 
\mathbf{x}\right) \right\rangle $ is implicitly included in the true shear
stress $\tau _{1}$.

Note that $\left\langle u^{\mathrm{down}}\left( \mathbf{x}\right)
\right\rangle _{A_{\mathrm{c}}}$ corresponds to the average contact plane. $%
\left\langle u^{\mathrm{down}}\left( \mathbf{x}\right) \right\rangle _{A_{%
\mathrm{c}}}$ is thus given by $\bar{u}^{\mathrm{down}}\left( \mathbf{x}%
\right) +\bar{u}_{\mathrm{cp}}^{\mathrm{down}}\left( \mathbf{x}\right) $,
where $\bar{u}_{\mathrm{cp}}^{\mathrm{down}}$ is the average separation
separating the average bottom solid surface $\bar{u}^{\mathrm{down}}\left( 
\mathbf{x}\right) $ from the average contact plane$\left\langle u^{\mathrm{%
down}}\left( \mathbf{x}\right) \right\rangle _{A_{\mathrm{c}}}$. Note that $%
\bar{u}_{\mathrm{cp}}^{\mathrm{down}}\left( \mathbf{x}\right) +\bar{u}_{%
\mathrm{cp}}^{\mathrm{up}}\left( \mathbf{x}\right) =\bar{u}\left( \mathbf{x}%
\right) $, where $\bar{u}_{\mathrm{cp}}^{\mathrm{up}}$ is the average
separation separating the average upper solid surface $\bar{u}^{\mathrm{up}%
}\left( \mathbf{x}\right) $ from the average contact plane$\left\langle u^{%
\mathrm{up}}\left( \mathbf{x}\right) \right\rangle _{A_{\mathrm{c}}}$, given
that $\left\langle u^{\mathrm{down}}\left( \mathbf{x}\right) \right\rangle
_{A_{\mathrm{c}}}=\left\langle u^{\mathrm{up}}\left( \mathbf{x}\right)
\right\rangle _{A_{\mathrm{c}}}$. To calculate the macro-rolling friction we
thus need to calculate $\bar{u}_{\mathrm{cp}}^{\mathrm{down}}\left( \mathbf{x%
}\right) $ (or $\bar{u}_{\mathrm{cp}}^{\mathrm{up}}\left( \mathbf{x}\right) $%
). We do this in an approximated way as follows. In particular, we consider
the variation $d\bar{u}_{\mathrm{cp}}^{\mathrm{down}}\left( \mathbf{x}%
\right) $ as due to a variation of the average applied solid contact
pressure $\sigma _{0}$%
\begin{equation*}
d\bar{u}_{\mathrm{cp}}^{\mathrm{down}}\left( \mathbf{x}\right) =\frac{1}{A_{%
\mathrm{c}}}\sum_{i}dA_{\mathrm{c}i}du_{i,\mathrm{cp}}^{\mathrm{down}},
\end{equation*}%
where we have adopted $\left\langle {}\right\rangle _{A_{\mathrm{c}}}=A_{%
\mathrm{c}}^{-1}\sum_{i}dA_{\mathrm{c}i}$ for homogeneous processes, and
where $dA_{\mathrm{c}i}$ is representative of the i-th asperity contact
area. Since $du_{i,\mathrm{cp}}^{\mathrm{down}}=-dw_{i}^{\mathrm{down}}$,
where $dw_{i}^{\mathrm{down}}$ is the local variation of displacement field,
we thus write%
\begin{equation*}
d\bar{u}_{\mathrm{cp}}^{\mathrm{down}}\left( \mathbf{x}\right) =-\frac{1}{A_{%
\mathrm{c}}}\sum_{i}dA_{\mathrm{c}i}dw_{i}^{\mathrm{down}}.
\end{equation*}%
By considering that, for practical applications, the average effective solid
pressure acting in the generic i-th asperity contact area $\sigma _{i}$ is
almost constantly valued (i.e. not dependent on the externally applied load,
unless for very large contact areas), we can write%
\begin{equation*}
\frac{1}{A_{\mathrm{c}}}\sum_{i}dA_{\mathrm{c}i}dw_{i}^{\mathrm{down}%
}\approx \frac{1}{A_{0}\sigma _{0}}\sum_{i}dA_{\mathrm{c}i}\sigma
_{i}dw_{i}^{\mathrm{down}}=\frac{1}{A_{0}\sigma _{0}}dU_{\mathrm{el}}^{%
\mathrm{down}}\left( \mathbf{x}\right) ,
\end{equation*}%
where $dU_{\mathrm{el}}^{\mathrm{down}}$ is the interfacial elastic energy
stored in the bottom solid. Therefore, the following differential relation
applies in the linear solid contact regime%
\begin{equation*}
d\bar{u}_{\mathrm{cp}}^{\mathrm{down}}A_{0}\sigma _{0}=-dU_{\mathrm{el}}^{%
\mathrm{down}}.
\end{equation*}%
Upon integrating the previous equation, and for the case of elastic contact
where only one of the solids is rough, one can relatively easily show that%
\begin{align*}
\bar{u}_{\mathrm{cp}}^{\mathrm{down}}\left( \mathbf{x}\right) & =\bar{u}%
\left( \mathbf{x}\right) \frac{E_{\mathrm{r}}^{\mathrm{up}}}{E_{\mathrm{r}}^{%
\mathrm{down}}+E_{\mathrm{r}}^{\mathrm{up}}} \\
\bar{u}_{\mathrm{cp}}^{\mathrm{up}}\left( \mathbf{x}\right) & =\bar{u}\left( 
\mathbf{x}\right) \frac{E_{\mathrm{r}}^{\mathrm{down}}}{E_{\mathrm{r}}^{%
\mathrm{down}}+E_{\mathrm{r}}^{\mathrm{up}}}.
\end{align*}%
Therefore, the macro-rolling solid friction term reads for the upper solid%
\begin{equation*}
-\bar{p}_{\mathrm{cont}}(x,t)\left[ \nabla \bar{u}^{\mathrm{up}%
}(x,t)-\nabla \bar{u}(x,t)\frac{E_{\mathrm{r}}^{\mathrm{down}}}{E_{\mathrm{r}%
}^{\mathrm{down}}+E_{\mathrm{r}}^{\mathrm{up}}}\right] 
\end{equation*}%
and for the lower solid%
\begin{equation*}
\bar{p}_{\mathrm{cont}}(x,t)\left[ \nabla \bar{u}^{\mathrm{down}%
}(x,t)+\nabla \bar{u}(x,t)\frac{E_{\mathrm{r}}^{\mathrm{up}}}{E_{\mathrm{r}%
}^{\mathrm{down}}+E_{\mathrm{r}}^{\mathrm{up}}}\right] ,
\end{equation*}%
where%
\begin{equation*}
\nabla \bar{u}^{\mathrm{up}}(x,t)-\nabla \bar{u}(x,t)\frac{E_{\mathrm{r}}^{%
\mathrm{down}}}{E_{\mathrm{r}}^{\mathrm{down}}+E_{\mathrm{r}}^{\mathrm{up}}}%
=\nabla \bar{u}^{\mathrm{down}}(x,t)+\nabla \bar{u}(x,t)\frac{E_{\mathrm{r}%
}^{\mathrm{up}}}{E_{\mathrm{r}}^{\mathrm{down}}+E_{\mathrm{r}}^{\mathrm{up}}}%
.
\end{equation*}

\section{}
In this section we describe how to calculate the frictional correction
factors for the generic contact case where the roughness is included only on
one of the two contact surfaces. The most general case where both surfaces
are rough has been described by some of us in a recent paper \cite{MS4,MS3}
by using a different homogenization procedure than used here. In a near
future the authors will provide a unified homogenization framework for the
most general wet contact case.

In Ref. \cite%
{MS2} we show how to calculate $\left\langle u\nabla p\right\rangle $ in the
low contact pressure regime $\bar{u}\gg h_{rms}$ and high contact pressure
regime $\bar{u}\ll \bar{u}_{c}$. We summarize here the main effective equations:

\textbf{For the case where the
rough surface is fixed} \cite{MS2}: 
\begin{eqnarray}
\bar{u}\gg h_{\mathrm{rms}} &:&\left\langle u\nabla p\right\rangle =a_{1}%
\bar{u}\nabla \bar{p}+a_{2}\frac{2\mathbf{v}_{0}\eta _{0}}{\bar{u}}
\label{A1} \\
\bar{u}\ll \bar{u}_{\mathrm{c}} &:&\left\langle u\nabla p\right\rangle =3%
\bar{u}\left\langle \frac{1}{u}\right\rangle \frac{2v_{0}\eta }{\bar{u}},
\label{A2}
\end{eqnarray}%
corresponding, respectively, to (16) and (17) and leading to an effective
flow 
\begin{equation*}
J_{0}=-\phi _{\mathrm{p}}\frac{\bar{u}^{3}}{12\eta }\nabla \bar{p}+\frac{1}{2%
}\phi _{\mathrm{s}}h_{\mathrm{rms}}v_{0}+\frac{1}{2}\bar{u}v_{0}.
\end{equation*}%

\textbf{For the case where the rough surface is sliding}: 

Using the same procedure as in Ref. \cite{MS2} one we obtain
\begin{eqnarray}
\bar{u}\gg h_{\mathrm{rms}} &:&\left\langle u\nabla p\right\rangle =a_{1}%
\bar{u}\nabla \bar{p}-a_{2}\frac{2\mathbf{v}_{0}\eta _{0}}{\bar{u}}
\label{A3} \\
\bar{u}\ll \bar{u}_{\mathrm{c}} &:&\left\langle u\nabla p\right\rangle =-3%
\bar{u}\left\langle \frac{1}{u}\right\rangle \frac{2v_{0}\eta }{\bar{u}},
\label{A4}
\end{eqnarray}%
leading to an effective flow 
\begin{equation*}
J_{0}=-\phi _{\mathrm{p}}\frac{\bar{u}^{3}}{12\eta }\nabla \bar{p}-\frac{1}{2%
}\phi _{\mathrm{s}}h_{\mathrm{rms}}v_{0}+\frac{1}{2}\bar{u}v_{0}.
\end{equation*}%
In (\ref{A1}) to (\ref{A4}) we have 
\begin{eqnarray*}
a_{1} &=&\left( 1-\langle h^{2}\rangle D{\frac{3}{\bar{u}^{2}}}\right)  \\
a_{2} &=&3D\frac{\langle h^{2}\rangle }{u^{2}}.
\end{eqnarray*}%
We now consider the effective frictional stresses. We write the separation,
at first order, as%
\begin{equation*}
u\left( \mathbf{x}\right) =\bar{u}\left( \mathbf{x}\right) +u_{1}^{\mathrm{up%
}}\left( \mathbf{x}\right) +u_{1}^{\mathrm{down}}\left( \mathbf{x}\right) ,
\end{equation*}%
where $u_{1}^{\mathrm{up}}\left( \mathbf{x}\right) =w_{1}^{\mathrm{up}%
}\left( \mathbf{x}\right) -h^{\mathrm{up}}\left( \mathbf{x}\right) $ and $%
u_{1}^{\mathrm{down}}\left( \mathbf{x}\right) =w_{1}^{\mathrm{down}}\left( 
\mathbf{x}\right) -h^{\mathrm{down}}\left( \mathbf{x}\right) $, where $%
w_{1}^{\mathrm{\cdot }}$ is the first order surface displacement field
(positive when inward the solid) and $h^{\mathrm{\cdot }}$ is the surface
roughness. Furthermore, we define the deformed locally averaged solids
profile as $\bar{u}^{\mathrm{up}}\left( \mathbf{x}\right) $ and $\bar{u}^{%
\mathrm{down}}\left( \mathbf{x}\right) $, with $\bar{u}\left( \mathbf{x}%
\right) =\bar{u}^{\mathrm{up}}\left( \mathbf{x}\right) -\bar{u}^{\mathrm{down%
}}\left( \mathbf{x}\right) $ (see also in the main text). In this work we
only consider the case where $h^{\mathrm{up}}\left( \mathbf{x}\right) =0$ or 
$h^{\mathrm{down}}\left( \mathbf{x}\right) =0$.\\
\textbf{a) Upper surface fixed - lower surface sliding}\newline
In this case, the frictional shear stress acting on the upper surface is%
\begin{equation*}
\tau_{\mathrm{fluid}} ^{\mathrm{up}}\left( \mathbf{x}\right) =\frac{\eta
v_{0}}{u\left( \mathbf{x}\right) }-\frac{1}{2}u\left( \mathbf{x}\right)
\nabla p\left( \mathbf{x}\right) -p\left( \mathbf{x}\right) \nabla \left( 
\bar{u}^{\mathrm{up}}\left( \mathbf{x}\right) +u_{1}^{\mathrm{up}}\left( 
\mathbf{x}\right) \right) ,
\end{equation*}%
whereas on the lower surface%
\begin{equation*}
\tau_{\mathrm{fluid}} ^{\mathrm{down}}\left( \mathbf{x}\right) =-\frac{\eta
v_{0}}{u\left( \mathbf{x}\right) }-\frac{1}{2}u\left( \mathbf{x}\right)
\nabla p\left( \mathbf{x}\right) +p\left( \mathbf{x}\right) \nabla \left( 
\bar{u}^{\mathrm{down}}\left( \mathbf{x}\right) -u_{1}^{\mathrm{down}}\left( 
\mathbf{x}\right) \right) .
\end{equation*}%
After some manipulation one gets%
\begin{eqnarray}
\left\langle \tau_{\mathrm{fluid}} ^{\mathrm{up}}\left( \mathbf{x}\right)
\right\rangle &=&\bar{\tau} _{\mathrm{visc}}^{\mathrm{up}}\left( \mathbf{x}%
\right) -\bar{p}\nabla \bar{u}^{\mathrm{up}}  \label{caseA.up} \\
\bar{\tau} _{\mathrm{visc}}^{\mathrm{up}}\left( \mathbf{x}\right) &=&\left(
v_{0}\left\langle \frac{\eta }{u}\right\rangle +\frac{1}{2}\left\langle
u\nabla p\right\rangle -\left\langle u_{1}^{\mathrm{down}}\nabla
p_{1}\right\rangle \right) -\bar{u}\nabla \bar{p}  \label{caseA.up1}
\end{eqnarray}%
and%
\begin{eqnarray}
\left\langle \tau_{\mathrm{fluid}} ^{\mathrm{down}}\left( \mathbf{x}\right)
\right\rangle &=&\bar{\tau} _{\mathrm{visc}}^{\mathrm{down}}\left( \mathbf{x}%
\right) +\bar{p}\nabla \bar{u}^{\mathrm{down}}  \label{caseA.down} \\
\bar{\tau} _{\mathrm{visc}}^{\mathrm{down}}\left( \mathbf{x}\right)
&=&-\left( v_{0}\left\langle \frac{\eta }{u}\right\rangle +\frac{1}{2}%
\left\langle u\nabla p\right\rangle -\left\langle u_{1}^{\mathrm{down}%
}\nabla p_{1}\right\rangle \right) .  \label{caseA.down1}
\end{eqnarray}%
In (\ref{caseA.up}) and (\ref{caseA.down}) $\bar{p}\nabla \bar{u}^{\mathrm{up%
}}$ and $\bar{p}\nabla \bar{u}^{\mathrm{down}}$ correspond to the fluid
rolling friction terms coming from the macroscopic deformed profile of the
solids. $\left\langle u_{1}^{\mathrm{down}}\nabla p_{1}\right\rangle $ can
be calculated by considering that, neglecting fluid-asperity flattening, 
\begin{eqnarray*}
\text{\textrm{for the}\textit{\ smooth}\textrm{\ bottom surface}}
&:&\left\langle u_{1}^{\mathrm{down}}\nabla p_{1}\right\rangle =0 \\
\text{\textrm{for the}\textit{\ rough}\textrm{\ bottom surface}}
&:&\left\langle u_{1}^{\mathrm{down}}\nabla p_{1}\right\rangle =\left\langle
u\nabla p\right\rangle -\bar{u}\nabla \bar{p}.
\end{eqnarray*}

\textit{a.1) Lower surface is smooth}\newline
In this case $\left\langle u_{1}^{\mathrm{down}}\nabla p_{1}\right\rangle =0$
resulting in 
\begin{equation*}
\bar{\tau}_{\mathrm{visc}}^{\mathrm{up}}\left( \mathbf{x}\right)
=v_{0}\left\langle \frac{\eta }{u}\right\rangle +\frac{1}{2}\left\langle
u\nabla p\right\rangle -\bar{u}\nabla \bar{p}
\end{equation*}%
and%
\begin{equation*}
\bar{\tau}_{\mathrm{visc}}^{\mathrm{down}}\left( \mathbf{x}\right) =-\left(
v_{0}\left\langle \frac{\eta }{u}\right\rangle +\frac{1}{2}\left\langle
u\nabla p\right\rangle \right) .
\end{equation*}%
We write, using (\ref{A1}) and (\ref{A2}) 
\begin{equation*}
\left\langle \frac{\eta }{u}\right\rangle =\phi _{f}\frac{\eta }{\bar{u}}
\end{equation*}%
and 
\begin{gather*}
\left\langle u\nabla p\right\rangle _{|\bar{u}\gg h_{\mathrm{rms}}}=\left(
1-\langle h^{2}\rangle D{\frac{3}{\bar{u}^{2}}}\right) \bar{u}\nabla \bar{p}%
+3D\frac{\langle h^{2}\rangle }{u^{2}}\frac{2\mathbf{v}_{0}\eta _{0}}{\bar{u}%
} \\
\left\langle u\nabla p\right\rangle _{|\bar{u}\ll \bar{u}_{\mathrm{c}}}=3%
\bar{u}\left\langle \frac{1}{u}\right\rangle \frac{2v_{0}\eta }{\bar{u}}
\end{gather*}%
leading, after interpolation, to 
\begin{equation*}
\left\langle u\nabla p\right\rangle =\phi _{\mathrm{fp}}\bar{u}\nabla \bar{p}%
+\phi _{\mathrm{fs}}\frac{2\mathbf{v}_{0}\eta _{0}}{\bar{u}},
\end{equation*}%
with 
\begin{eqnarray*}
\phi _{\mathrm{fp}} &=&{\frac{\bar{u}(\bar{u}-\bar{u}_{\mathrm{c}})\theta (%
\bar{u}-\bar{u}_{\mathrm{c}})}{\bar{u}^{2}+3\langle h^{2}\rangle D}} \\
\phi _{\mathrm{fs}} &=&\frac{3\bar{u}}{\langle u^{-1}\rangle ^{-1}+\theta (%
\bar{u}-\bar{u}_{\mathrm{c}})(\bar{u}-\bar{u}_{\mathrm{c}})(\bar{u}%
^{2}/\langle h^{2}\rangle )D^{-1}}.
\end{eqnarray*}%
Finally we have 
\begin{eqnarray*}
\bar{\tau}_{\mathrm{visc}}^{\mathrm{up}} &=&\left( \phi _{\mathrm{f}}+\phi _{%
\mathrm{fs}}\right) {\frac{\eta _{0}\mathbf{v}_{0}}{\bar{u}}}+\frac{1}{2}%
\left( \phi _{\mathrm{fp}}-2\right) \bar{u}\nabla \bar{p} \\
\bar{\tau}_{\mathrm{visc}}^{\mathrm{down}} &=&-\left( \phi _{\mathrm{f}%
}+\phi _{\mathrm{fs}}\right) {\frac{\eta _{0}\mathbf{v}_{0}}{\bar{u}}}-\frac{%
1}{2}\phi _{\mathrm{fp}}\bar{u}\nabla \bar{p}.
\end{eqnarray*}

\textit{a.2) Lower surface is rough}\newline
In that case $\left\langle u_{1}^{\mathrm{down}}\nabla p_{1}\right\rangle
=\left\langle u\nabla p\right\rangle -\bar{u}\nabla \bar{p}$ and%
\begin{equation*}
\tau _{\mathrm{visc}}^{\mathrm{up}}\left( \mathbf{x}\right)
=v_{0}\left\langle \frac{\eta }{u}\right\rangle -\frac{1}{2}\left\langle
u\nabla p\right\rangle 
\end{equation*}%
and%
\begin{equation*}
\tau _{\mathrm{visc}}^{\mathrm{down}}\left( \mathbf{x}\right) =-\left(
v_{0}\left\langle \frac{\eta }{u}\right\rangle -\frac{1}{2}\left\langle
u\nabla p\right\rangle +\bar{u}\nabla \bar{p}\right) .
\end{equation*}%
We write, using (\ref{A3}) and (\ref{A4})%
\begin{equation*}
\left\langle \frac{\eta }{u}\right\rangle =\phi _{\mathrm{f}}\frac{\eta }{%
\bar{u}}
\end{equation*}%
and%
\begin{gather*}
\left\langle u\nabla p\right\rangle _{|\bar{u}\gg h_{rms}}=\left( 1-\langle
h^{2}\rangle D{\frac{3}{\bar{u}^{2}}}\right) \bar{u}\nabla \bar{p}-3D\frac{%
\langle h^{2}\rangle }{u^{2}}\frac{2\mathbf{v}_{0}\eta _{0}}{\bar{u}} \\
\left\langle u\nabla p\right\rangle _{|\bar{u}\ll \bar{u}_{c}}=-3\bar{u}%
\left\langle \frac{1}{u}\right\rangle \frac{2v_{0}\eta }{\bar{u}},
\end{gather*}%
leading, after interpolation, to%
\begin{equation*}
\left\langle u\nabla p\right\rangle =\phi _{\mathrm{fp}}\bar{u}\nabla \bar{p}%
-\phi _{\mathrm{fs}}\frac{2\mathbf{v}_{0}\eta _{0}}{\bar{u}},
\end{equation*}%
where $\phi _{\mathrm{fp}}$ and $\phi _{\mathrm{fs}}$ are as before. Finally
we have%
\begin{eqnarray*}
\tau _{\mathrm{visc}}^{\mathrm{up}} &=&\left( \phi _{\mathrm{f}}+\phi _{%
\mathrm{fs}}\right) {\frac{\eta _{0}\mathbf{v}_{0}}{\bar{u}}}-\frac{1}{2}%
\phi _{\mathrm{fp}}\bar{u}\nabla \bar{p} \\
\tau _{\mathrm{visc}}^{\mathrm{down}} &=&-\left( \phi _{\mathrm{f}}+\phi _{%
\mathrm{fs}}\right) {\frac{\eta _{0}\mathbf{v}_{0}}{\bar{u}}}+\frac{1}{2}%
\left( \phi _{\mathrm{fp}}-2\right) \bar{u}\nabla \bar{p}.
\end{eqnarray*}%
Summarizing:%
\begin{equation*}
\begin{tabular}{|c|c|c|}
\hline
& Up fixed & Low fixed \\ \hline
Up rough & $%
\begin{array}{c}
\bar{\tau}_{\mathrm{visc}}^{\mathrm{up}}=\left( \phi _{\mathrm{f}}+\phi _{%
\mathrm{fs}}\right) {\frac{\eta _{0}\mathbf{v}_{0}}{\bar{u}}}+\frac{1}{2}%
\left( \phi _{\mathrm{fp}}-2\right) \bar{u}\nabla \bar{p} \\ 
\bar{\tau}_{\mathrm{visc}}^{\mathrm{down}}=-\left( \phi _{\mathrm{f}}+\phi _{%
\mathrm{fs}}\right) {\frac{\eta _{0}\mathbf{v}_{0}}{\bar{u}}}-\frac{1}{2}%
\phi _{\mathrm{fp}}\bar{u}\nabla \bar{p}%
\end{array}%
$ & $%
\begin{array}{c}
\bar{\tau}_{\mathrm{visc}}^{\mathrm{up}}=-\left( \phi _{\mathrm{f}}+\phi _{%
\mathrm{fs}}\right) {\frac{\eta _{0}\mathbf{v}_{0}}{\bar{u}}}+\frac{1}{2}%
\left( \phi _{\mathrm{fp}}-2\right) \bar{u}\nabla \bar{p} \\ 
\bar{\tau}_{\mathrm{visc}}^{\mathrm{down}}=\left( \phi _{\mathrm{f}}+\phi _{%
\mathrm{fs}}\right) {\frac{\eta _{0}\mathbf{v}_{0}}{\bar{u}}}-\frac{1}{2}%
\phi _{\mathrm{fp}}\bar{u}\nabla \bar{p}%
\end{array}%
$ \\ \hline
Low rough & $%
\begin{array}{c}
\tau _{\mathrm{visc}}^{\mathrm{up}}=\left( \phi _{\mathrm{f}}+\phi _{\mathrm{%
fs}}\right) {\frac{\eta _{0}\mathbf{v}_{0}}{\bar{u}}}-\frac{1}{2}\phi _{%
\mathrm{fp}}\bar{u}\nabla \bar{p} \\ 
\tau _{\mathrm{visc}}^{\mathrm{down}}=-\left( \phi _{\mathrm{f}}+\phi _{%
\mathrm{fs}}\right) {\frac{\eta _{0}\mathbf{v}_{0}}{\bar{u}}}+\frac{1}{2}%
\left( \phi _{\mathrm{fp}}-2\right) \bar{u}\nabla \bar{p}%
\end{array}%
$ & $%
\begin{array}{c}
\bar{\tau}_{\mathrm{visc}}^{\mathrm{up}}=-\left( \phi _{\mathrm{f}}+\phi _{%
\mathrm{fs}}\right) {\frac{\eta _{0}\mathbf{v}_{0}}{\bar{u}}}-\frac{1}{2}%
\phi _{\mathrm{fp}}\bar{u}\nabla \bar{p} \\ 
\bar{\tau}_{\mathrm{visc}}^{\mathrm{down}}=\left( \phi _{\mathrm{f}}+\phi _{%
\mathrm{fs}}\right) {\frac{\eta _{0}\mathbf{v}_{0}}{\bar{u}}}+\frac{1}{2}%
\left( \phi _{\mathrm{fp}}-2\right) \bar{u}\nabla \bar{p}%
\end{array}%
$ \\ \hline
\end{tabular}%
\end{equation*}%
\newline

\end{widetext}


\bibliographystyle{unsrt}
\bibliography{articlebib}

\begin{thebibliography}{10}

\bibitem{general_multiscale}
Scaraggi M. and Persson~Bo N.J.
\newblock General contact mechanics theory for randomly rough surfaces with
  application to rubber friction.
\newblock {\em The Journal of Chemical Physics}, 2015.

\bibitem{Ma20157366}
S.~Ma, M.~Scaraggi, D.~Wang, X.~Wang, Y.~Liang, W.~Liu, D.~Dini, and F.~Zhou.
\newblock Nanoporous substrate-infiltrated hydrogels: A bioinspired regenerable
  surface for high load bearing and tunable friction.
\newblock {\em Advanced Functional Materials}, 25(47):7366--7374, 2015.

\bibitem{10.1371/journal.pone.0143415}
Edward~D. Bonnevie, Devis Galesso, Cynthia Secchieri, Itai Cohen, and
  Lawrence~J. Bonassar.
\newblock Elastoviscous transitions of articular cartilage reveal a mechanism
  of synergy between lubricin and hyaluronic acid.
\newblock {\em PLoS ONE}, 10(11):1--15, 11 2015.

\bibitem{Stupkiewicz2016511}
S.~Stupkiewicz, J.~Lengiewicz, P.~Sadowski, and S.~Kucharski.
\newblock Finite deformation effects in soft elastohydrodynamic lubrication
  problems.
\newblock {\em Tribology International}, 93:511--522, 2016.

\bibitem{Salant1999189}
R.F. Salant.
\newblock Theory of lubrication of elastomeric rotary shaft seals.
\newblock {\em Proceedings of the Institution of Mechanical Engineers, Part J:
  Journal of Engineering Tribology}, 213(3):189--201, 1999.

\bibitem{Hajjam200413}
M.~Hajjam and D.~Bonneau.
\newblock Elastohydrodynamic analysis of lip seals with microundulations.
\newblock {\em Proceedings of the Institution of Mechanical Engineers, Part J:
  Journal of Engineering Tribology}, 218(1):13--21, 2004.

\bibitem{Wohlers200951}
A.~Wohlers, O.~Heipl, B.N.J. Persson, M.~Scaraggi, and H.~Murrenhoff.
\newblock Numerical and experimental investigation on o-ring-seals in dynamic
  applications.
\newblock {\em International Journal of Fluid Power}, 10(3):51--59, 2009.

\bibitem{Scaraggi2012}
M.~Scaraggi and G.~Carbone.
\newblock A two-scale approach for lubricated soft-contact modeling: An
  application to lip-seal geometry.
\newblock {\em Advances in Tribology}, 2012.

\bibitem{Tan2015236}
Gui-Bin Tan, Shu-Hai Liu, De-Guo Wang, and Si-Wei Zhang.
\newblock Spatio-temporal structure in wax–oil gel scraping at a soft
  tribological contact.
\newblock {\em Tribology International}, 88:236 -- 251, 2015.

\bibitem{Scaraggi2014118}
M.~Scaraggi and B.N.J. Persson.
\newblock Theory of viscoelastic lubrication.
\newblock {\em Tribology International}, 72:118--130, 2014.

\bibitem{Dunn201345}
A.C. Dunn, J.A. Tichy, J.M. Uruenã, and W.G. Sawyer.
\newblock Lubrication regimes in contact lens wear during a blink.
\newblock {\em Tribology International}, 63:45--50, 2013.

\bibitem{Khosla201517587}
T.~Khosla, J.~Cremaldi, J.S. Erickson, and N.S. Pesika.
\newblock Load-induced hydrodynamic lubrication of porous films.
\newblock {\em ACS Applied Materials and Interfaces}, 7(32):17587--17591, 2015.

\bibitem{Greene20115255}
G.W. Greene, X.~Banquy, D.~Woog~Lee, D.D. Lowrey, J.~Yu, and J.N.
  Israelachvili.
\newblock Adaptive mechanically controlled lubrication mechanism found in
  articular joints.
\newblock {\em Proceedings of the National Academy of Sciences of the United
  States of America}, 108(13):5255--5259, 2011.

\bibitem{Lorenz2013}
B.~Lorenz, B.A. Krick, N.~Rodriguez, W.G. Sawyer, P.~Mangiagalli, and B.N.J.
  Persson.
\newblock Static or breakloose friction for lubricated contacts: The role of
  surface roughness and dewetting.
\newblock {\em Journal of Physics Condensed Matter}, 25(44), 2013.

\bibitem{Sterner2016}
O.~Sterner, R.~Aeschlimann, S.~Zürcher, C.~Scales, D.~Riederer, N.D. Spencer,
  and S.G.P. Tosatti.
\newblock Tribological classification of contact lenses: From coefficient of
  friction to sliding work.
\newblock {\em Tribology Letters}, 63(1), 2016.

\bibitem{EU.tires}
Regulation (ec) no 1222/2009 of the {European} {Parliament} and of the
  {Council} of 25 {November} 2009.

\bibitem{pavement}
Ahmed Elghriany, Ping Yi, Peng Liu, and Quan Yu.
\newblock Investigation of the effect of pavement roughness on crash rates for
  rigid pavement.
\newblock {\em Journal of Transportation Safety \& Security}, 8(2):164--176,
  2016.

\bibitem{Flitney}
R.~Flitney.
\newblock {\em Seals and Sealing Handbook: Sixth Edition}.
\newblock 2014.
\newblock cited By 0.

\bibitem{Persson20013840}
Persson B.N.J.
\newblock Theory of rubber friction and contact mechanics.
\newblock {\em Journal of Chemical Physics}, 115(8):3840--3861, 2001.

\bibitem{Carbone2009}
Carbone G., Lorenz B., Persson B.N.J., and Wohlers A.
\newblock Contact mechanics and rubber friction for randomly rough surfaces
  with anisotropic statistical properties.
\newblock {\em The European Physical Journal E: Soft Matter and Biological
  Physics}, 29(3):275--284, 2009.

\bibitem{BP}
B.N.J. Persson.
\newblock Fluid dynamics at the interface between contacting elastic solids
  with randomly rough surfaces.
\newblock {\em Journal of Physics Condensed Matter}, 22(26), 2010.
\newblock cited By 23.

\bibitem{MS1}
B.N.J. Persson and M.~Scaraggi.
\newblock On the transition from boundary lubrication to hydrodynamic
  lubrication in soft contacts.
\newblock {\em Journal of Physics Condensed Matter}, 21(18), 2009.
\newblock cited By 39.

\bibitem{MS2}
B.N.J. Persson and M.~Scaraggi.
\newblock Lubricated sliding dynamics: Flow factors and stribeck curve.
\newblock {\em European Physical Journal E}, 34(10):113, 2011.
\newblock cited By 17.

\bibitem{MS3}
M.~Scaraggi, G.~Carbone, and D.~Dini.
\newblock {Lubrication in soft rough contacts: A novel homogenized approach.
  Part II - Discussion}.
\newblock {\em Soft Matter}, 7(21):10407--10416, 2011.
\newblock cited By 19.

\bibitem{MS4}
M.~Scaraggi, G.~Carbone, B.N.J. Persson, and D.~Dini.
\newblock {Lubrication in soft rough contacts: A novel homogenized approach.
  Part I - Theory}.
\newblock {\em Soft Matter}, 7(21):10395--10406, 2011.
\newblock cited By 34.

\bibitem{MS5}
M.~Scaraggi.
\newblock Lubrication of textured surfaces: A general theory for flow and shear
  stress factors.
\newblock {\em Physical Review E - Statistical, Nonlinear, and Soft Matter
  Physics}, 86(2), 2012.
\newblock cited By 16.

\bibitem{MS6}
M.~Scaraggi and B.N.J. Persson.
\newblock Time-dependent fluid squeeze-out between soft elastic solids with
  randomly rough surfaces.
\newblock {\em Tribology Letters}, 47(3):409--416, 2012.
\newblock cited By 10.

\bibitem{Patir197812}
N.~Patir and H.S. Cheng.
\newblock An average flow model for determining effects of three-dimensional
  roughness on partial hydrodynamic lubrication.
\newblock {\em Trans. ASME Ser. F, J. Lubr. Technol.}, 100(1 , Jan.
  1978):12--17, 1978.
\newblock cited By 946.

\bibitem{Persson2007}
B.N.J. Persson.
\newblock Relation between interfacial separation and load: A general theory of
  contact mechanics.
\newblock {\em Physical Review Letters}, 99(12), 2007.
\newblock cited By 77.

\bibitem{PS0}
Persson B.N.J.
\newblock ....
\newblock {\em Journal of ...}, 27(10):105102, 2015.

\bibitem{Persson2005}
Persson B.N.J., Albohr O., Tartaglino U., Volokitin A.I., and Tosatti E.
\newblock On the nature of surface roughness with application to contact
  mechanics, sealing, rubber friction and adhesion.
\newblock {\em Journal of Physics Condensed Matter}, 17(1):R1--R62, 2005.

\bibitem{MS7}
M.~Scaraggi, G.~Carbone, and D.~Dini.
\newblock Experimental evidence of micro-ehl lubrication in rough soft
  contacts.
\newblock {\em Tribology Letters}, 43(2):169--174, 2011.
\newblock cited By 19.

\bibitem{Scaraggi.in.prep}
Scaraggi M. and Persson B.N.J.
\newblock Friction and universal contact area law for randomly rough
  viscoelastic contacts.
\newblock {\em Journal of Physics Condensed Matter}, 27(10), 2015.

\bibitem{scaraggi2016effect}
M~Scaraggi and BNJ Persson.
\newblock The effect of finite roughness size and bulk thickness on the
  prediction of rubber friction and contact mechanics.
\newblock {\em Proceedings of the Institution of Mechanical Engineers, Part C:
  Journal of Mechanical Engineering Science}, 230(9):1398--1409, 2016.

\bibitem{lorenz2010leak}
B~Lorenz and B~NJ Persson.
\newblock Leak rate of seals: Effective-medium theory and comparison with
  experiment.
\newblock {\em The European Physical Journal E: Soft Matter and Biological
  Physics}, 31(2):159--167, 2010.

\bibitem{lorenz2013adhesion}
B~Lorenz, BA~Krick, N~Mulakaluri, M~Smolyakova, S~Dieluweit, WG~Sawyer, and BNJ
  Persson.
\newblock Adhesion: role of bulk viscoelasticity and surface roughness.
\newblock {\em Journal of Physics: Condensed Matter}, 25(22):225004, 2013.

\bibitem{lorenz2013static}
B~Lorenz, BA~Krick, N~Rodriguez, WG~Sawyer, P~Mangiagalli, and BNJ Persson.
\newblock Static or breakloose friction for lubricated contacts: the role of
  surface roughness and dewetting.
\newblock {\em Journal of Physics: Condensed Matter}, 25(44):445013, 2013.

\bibitem{Persson2014}
Persson B.N.J. and Scaraggi M.
\newblock Friction and universal contact area law for randomly rough
  viscoelastic contacts.
\newblock {\em Journal of Physics: Condensed Matter}, 27(10):105102, 2015.

\bibitem{0953-8984-21-1-015003}
B~Lorenz and B~N~J Persson.
\newblock Interfacial separation between elastic solids with randomly rough
  surfaces: comparison of experiment with theory.
\newblock {\em Journal of Physics: Condensed Matter}, 21(1):015003, 2009.

\bibitem{Almqvist20112355}
Almqvist A., Campa{\~n}{\'a} C., Prodanov N., and Persson B.N.J.
\newblock Interfacial separation between elastic solids with randomly rough
  surfaces: Comparison between theory and numerical techniques.
\newblock {\em Journal of the Mechanics and Physics of Solids},
  59(11):2355--2369, 2011.

\bibitem{Greenwood300}
J.~A. Greenwood and J.~B.~P. Williamson.
\newblock Contact of nominally flat surfaces.
\newblock {\em Proceedings of the Royal Society of London A: Mathematical,
  Physical and Engineering Sciences}, 295(1442):300--319, 1966.

\bibitem{Khonsari}
M~Masjedi and MM~Khonsari.
\newblock Mixed lubrication of soft contacts: An engineering look.
\newblock {\em Proceedings of the Institution of Mechanical Engineers, Part J:
  Journal of Engineering Tribology}, 231(2):263--273, 2017.

\bibitem{nayak}
R.P. Nayak.
\newblock Random process model of rough surfaces.
\newblock {\em J Lubric Technol Trans ASME}, 93 Ser F(3):398--407, 1971.

\bibitem{conf.23}
J.~Angerhausen, H.~Murrenhoff, L.~Dorogin, M.~Scaraggi, B.~Lorenz, and Bo~N.J.
  Persson.
\newblock Influence of anisotropic surfaces on the friction behaviour of
  hydraulic seals.
\newblock In {\em Proceedings of the 2016 Bath/ASME Symposium on Fluid Power
  and Motion Control}, 2016.

\end{thebibliography}

\end{document}